\documentclass[fleqn,10pt]{wlscirep}
\usepackage[utf8]{inputenc}
\usepackage[T1]{fontenc}
\usepackage{placeins}
\usepackage{float}
\usepackage{caption}
\usepackage{subfig}
\usepackage{graphicx}
\usepackage{wrapfig}
\usepackage{tabularx}
\usepackage{hyperref}
\usepackage{pdfpages}

\makeatletter
  \renewcommand\thesubfigure{\Alph{subfigure}}
  \renewcommand\p@subfigure{}       
\makeatother

\title{Machine Learning for analysis of Multiple Sclerosis cross-tissue bulk and single-cell transcriptomics data}

\author[1,+]{Francesco Massafra}
\author[1,+]{Samuele Punzo}
\author[1,*]{Silvia Giulia Galfré}
\author[2]{Alessandro Maglione}
\author[2,3]{Simone Pernice}
\author[1] {Stefano Forti}
\author[4]{Simona Rolla}
\author[2,3]{Marco Beccuti}
\author[4]{Marinella Clerico}
\author[1]{Corrado Priami}
\author[5]{Alina  S\^irbu}
\affil[1]{Department of Computer Science, University of Pisa, Pisa, Italy.}
\affil[2]{Department of Computer Science, University of Turin, Turin, Italy}
\affil[3]{ Laboratorio InfoLife, Consorzio Interuniversitario Nazionale per l'Informatica, Rome, Italy}
\affil[4]{Department of Clinical and Biological Sciences, University of Turin, Turin, Italy}
\affil[5]{Department of Computer Science and Engineering, University of Bologna, Bologna, Italy.}

\affil[*]{silvia.galfre@unipi.it}

\affil[+]{these authors contributed equally to this work}


\begin{abstract}

Multiple Sclerosis (MS) is a chronic autoimmune disease of the central nervous system whose molecular mechanisms remain incompletely understood. In this study, we developed an end-to-end machine learning pipeline to analyze transcriptomic data from peripheral blood mononuclear cells  and cerebrospinal fluid, integrating both bulk microarray and single-cell RNA sequencing datasets (concentrating on CD4+ and B-cells). After rigorous preprocessing, batch correction, and gene declustering, XGBoost classifiers were trained to distinguish MS patients from healthy controls. Explainable AI tools, namely SHapley Additive exPlanations (SHAP), were employed to identify key genes driving classification, and results were compared with Differential Expression Analysis (DEA). SHAP-prioritized genes were further investigated through interaction networks and pathway enrichment analyses. The models achieved strong performance, particularly in CSF B-cells (AUC = 0.94) and microarray (AUC=0.86). SHAP gene selection proved to be complementary to classical DEA. Gene clusters identified across multiple datasets highlighted immune activation, non-canonical immune checkpoints (ITK, CLEC2D, KLRG1, CEACAM1), ribosomal and translational programs, ubiquitin–proteasome regulation, lipid trafficking, and Epstein–Barr virus-related pathways. Our integrative and explainable framework reveals complementary insights beyond conventional analysis and provides novel mechanistic hypotheses and potential biomarkers for MS pathogenesis.
\end{abstract}

\begin{document}

\flushbottom
\maketitle
%
%
\thispagestyle{empty}

\section*{Introduction}
\label{sec:Introduction}
Multiple Sclerosis (MS) is a chronic autoimmune disease of the central nervous system characterized by inflammation, demyelination, and neurodegeneration, which is estimated to affect 2.8 million people worldwide. In countries with the highest prevalence, it reaches the ratio of 1 in every 300 people having MS \cite{atlasMS}. Despite advances in understanding its pathogenesis, the molecular mechanisms underlying MS are still unknown. Some research uncovered the role of B-cells, CD4+ autoreactive T cells, along with their differentiation and some innate immune mechanisms\cite{MSpathogenesis}, but we still lack reliable biomarkers. 

Recent studies have shown the potential of Machine Learning (ML) methods to discover novel biomarkers by integrating multi-omics data\cite{MLomics}. Yet, the problem of interpreting model decisions and linking them to biological knowledge is still an open one. Explainable AI (xAI) techniques such as SHapley Additive exPlanations (SHAP)\cite{SHAP} offer a solution by quantifying feature importance in ML models, improving their transparency. 

In this study, we present an end-to-end pipeline to analyze publicly available scRNA-seq and Microarray datasets of MS patients and healthy controls, including Peripheral Blood Mononuclear Cells (PBMCs) and CerebroSpinal Fluid (CSF) samples. Our workflow includes rigorous preprocessing, batch correction, and declustering to mitigate technical biases, followed by XGBoost modeling\cite{XGB} to classify patients into MS and Control.  We employed SHAP-based xAI to identify key genes, compared findings with standard Differential Expression Analysis (DEA), and validated them through interaction networks and pathway enrichment. Our results highlight the complexity of neuroinflammation in MS, identifying pathways and clusters ranging from non-canonical immune checkpoints to lipid trafficking, protein post-translational modifications, and Epstein-Barr Virus (EBV) infection. By combining ML-driven feature selection with biological interpretation, this study provides a framework for uncovering novel mechanistic insights into MS and other complex diseases.

\section{Methods}
\subsection{Data}
Our study integrates two transcriptomic modalities, single-cell RNA sequencing (scRNA-seq) and microarrays, across two tissue types: peripheral blood mononuclear cells (PBMCs) and cerebrospinal fluid (CSF).

We selected two scRNA-seq datasets from the NCBI Gene Expression Omnibus (GEO): GSE138266~\cite{GSE138266} and GSE194078~\cite{GSE194078}. Both include PBMC and CSF samples from patients with multiple sclerosis (MS) and healthy controls. Because GSE194078 also contains individuals with neurological conditions other than MS, we filtered this dataset to retain only MS and control samples. We then constructed two tissue-specific datasets, each comprising all samples from a single tissue (PBMCs or CSF).

For microarray profiling, we included eight publicly available GEO PBMC datasets (GSE41848~\cite{GSE41848-9}, GSE41849~\cite{GSE41848-9}, GSE146383~\cite{GSE146383}, GSE13732~\cite{GSE13732}, GSE136411~\cite{GSE136411}, GSE17048~\cite{GSE17048}, GSE41890~\cite{GSE41890}, GSE21942~\cite{GSE21942}). Samples were annotated as either MS or healthy control, depending on the study design. All selected datasets comprised adult participants, and several included longitudinal follow-up samples from the same individuals.

Across both scRNA-seq and microarray datasets, when MS samples spanned multiple disease stages, we grouped them under a single MS label because stage-specific stratification was outside the scope of this study. In addition, none of the MS participants had received therapy at the time of sampling. We initially screened additional datasets (GSE61240, GSE59867, GSE19301, GSE88794 and GSE144744) but excluded them due to unbalanced study design (MS-only or control-only cohorts), which limits batch correction, or because substantial technical noise compromised data quality.

\begin{figure}[h]
\vspace*{0pt}
    \centering
     \includegraphics[width=0.9\linewidth]{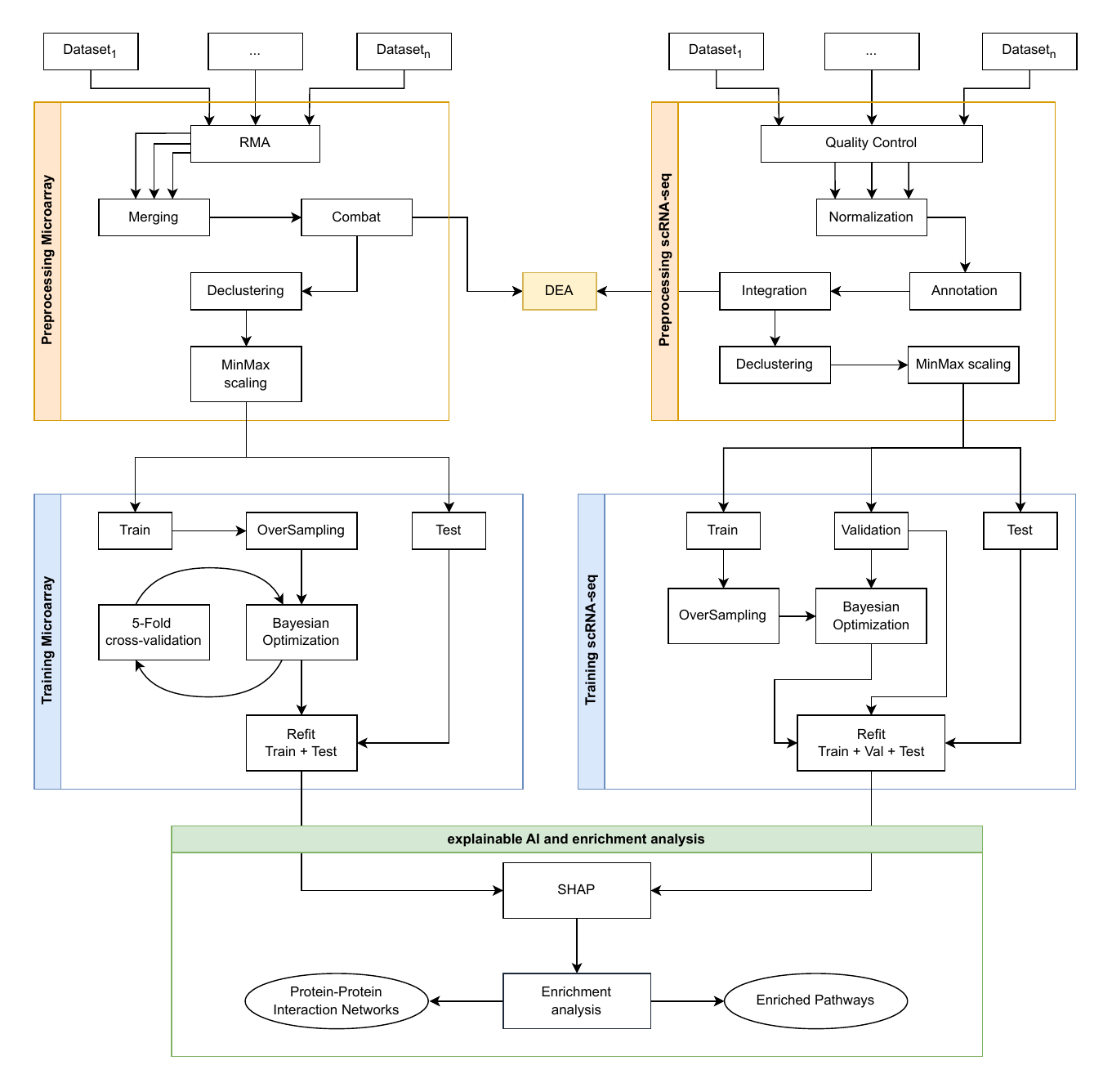}
     \captionof{figure}{Illustration of the implemented pipeline.}
     \label{fig:Pipeline}
\end{figure}

\subsection{ML analysis}
The implemented pipeline is shown in \autoref{fig:Pipeline} and can be split into: Preprocessing, Training, and Explainable AI (xAI). Differential Expression Analysis (DEA) was also performed to compare results with those from the xAI study. The analyses were performed in R and Python; both data and code are publicly available on \href{https://github.com/seriph78/ML_for_MS.git}{GitHub}.

\subsubsection{Preprocessing}
\label{ssec:Preprocessing}

\paragraph{scRNA-seq}
Prior to dataset integration, we performed Quality Control on each sample separately to remove low-quality cells and genes with low expression.
The Scanpy Python Package\cite{SCANPY} was used to filter cells that did not meet the criteria chosen for each dataset: number of genes expressed in each cell, number of Unique Molecular Identifiers (UMIs), and percentage of mitochondrial genes expressed.
Following quality control, each sample underwent normalization to a total cell count of 10,000 and was $\text{log}_{2}\text{-normalized}$. This step was performed to ensure that variance was comparable across samples and to reduce the impact of outliers \cite{cite-key}.
To facilitate robust integration, we used CellTypist\cite{CellTypist} to annotate cell types within each sample. We then generalized these annotations to establish a common set of cell types across all samples.
SCGen integration\cite{SCGEN} was then applied separately to samples from each tissue to effectively reduce underlying batch effects. This step resulted in two integrated datasets: one for PBMC and one for CSF, each comprising all cells and common genes across their respective samples.
To evaluate the quality of the preprocessing and integration steps, we used visual assessment via UMAP. 
To assess how distinct immune cell types contribute to MS pathology, we extracted two populations involved in MS: CD4+ T cells and B cells. We excluded naive cells from both populations and removed cell populations with insufficient numbers for robust downstream analysis. This yielded four scRNA-seq datasets in total: two derived from PBMCs and two from CSF.

\paragraph{Microarray}
We harmonized the microarray datasets via RMA normalization\cite{RMA}, a technique specific to microarrays that aligns intensity signal distributions by conducting background correction, quantile normalization, probe summarization, and $log_2$ transformation. This method was applied to each dataset individually, ensuring comparability within a single dataset. To address batch effects, we applied Combat\cite{Combat}, a method that reduces background noise by using Empirical Bayes methods to eliminate variability caused by non-biological factors. Finally, MinMax scaling was used to adjust the variable ranges to $[0, 1]$, ensuring that no variable would dominate the others during the training phase.

To evaluate preprocessing quality, both qualitative and quantitative techniques were employed, including: DEA between datasets under identical conditions, visual assessment of sample distribution via Principal Component Analysis (PCA), analysis of multimodal distributions, and Mixture Score quantification~\cite{MixtureScore} (a quantitative measure to evaluate the quality of batch effect removal). Together, these techniques provided a solid assessment of pre-processing quality and ensured data reliability for the following phases.

\paragraph{Declustering}
Before training the ML models, we performed for both data types a correlation analysis to identify clusters of genes with an absolute correlation higher than $0.9 \; (Pearson \; correlation)$. Each cluster was then substituted with a representative, i.e. the gene with the highest variance in the cluster. This phase is crucial to prevent certain genes from being overlooked as non-significant during the Explainable AI  (xAI) phase due to their biological similarity.

\subsubsection{Training}
\label{chap:training}
For the microarray data, we divided the samples into training and test ($75/25 \% $), ensuring that all follow-ups from the same subject were kept together. We obtained 510 Control and 262 MS Samples in training and
166 Control and 100 MS Samples in test.  The training dataset was used for training and model selection via 5-fold cross-validation, while the test dataset was used to evaluate the final model.  

For the scRNA-seq data, where multiple cells originate from the same individual, we split each dataset into three subsets: training, validation, and test ($60/20/20 \%$), grouping all cells from the same patient together, so that the same patient does not appear in both training and test/validation.  We opted for a single validation dataset rather than using 5-fold cross-validation because single-cell datasets contain multiple cells belonging to a small number of patients, which needed to be carefully grouped manually to prevent information leakage between sets. We include the number of MS and Control samples for each train/test/validation subset in \autoref{tab:divisione_training_val_test_singlecell_fixed} of the supplementary material.

In both types of data, we observed a severe imbalance between the two classes (especially in scRNA-seq datasets), with the MS class being significantly more represented than the Control class. To address this imbalance, techniques such as oversampling (SMOTE\cite{SMOTE}) and undersampling (Random) were applied to balance the classes in the training set by synthetically generating new samples for the minority class (Control) and randomly reducing the number of samples for the majority class (MS).

For each of the 5 datasets (4 scRNA-seq and 1 microarray), an XGBoost classifier\cite{XGB} was trained to distinguish between MS and Control samples/cells.  We used Bayesian hyperparameter optimization\cite{BayesianSearch} for model selection, a technique that efficiently searches the hyperparameter space (presented in \autoref{tab:scRNAseqHyperparameters} and \autoref{tab:microarrayHyperparameters} in the supplementary material). 
The best model was chosen based on the highest F1-score macro average on the validation set (or the mean 5-fold score for microarray data). We selected the F1-score as the evaluation metric to balance both precision and recall, ensuring relevant predictions for the minority class (Control). 
Finally, the best models were evaluated on their respective hold-out test sets, using both F1-score and the Area Under the ROC Curve (AUC).

\subsubsection{Explainable AI \& Enrichment analysis}
For each dataset, we performed an xAI analysis on our best ML model, which was retrained on the whole set of samples for each dataset (train, validation, and test).
To interpret the models' decision-making processes and extract biological insights, we employed SHAP\cite{SHAP} as a framework for feature importance analysis. In particular, the TreeExplainer~\cite{TreeExplainer} was used to rank features by their contribution to the model's predictions. Afterwards, the first $1,000$ features with the highest mean absolute SHAP values were selected. The genes related to the selected genes that were removed in the declustering step were added back, with all genes within a cluster assigned the same importance as the representative gene. For important genes, dependence plots were generated to study the relation between gene importance and gene expression values. 

These sets of genes were then used as a knowledge base to perform gene enrichment analysis using StringDB\cite{StringDB}, KEGG\cite{Kegg}, and Reactome\cite{Reactome} databases, allowing us to identify potential pathways and biological relations between genes. First, PPI (Protein-Protein Interaction Networks) were built using StringDB, with interactions filtered at a confidence threshold of $0.400$ to ensure high reliability\cite{StringScore}. From these networks, an overrepresentation analysis (ORA) was performed via StringDB, resulting in several significantly enriched pathways (adj. \textit{p}-value $< 0.05$) for each dataset from both KEGG and Reactome databases. 
Enriched pathways were further examined for biological coherence, with particular attention to those implicated in the disease or other interesting conditions.


\subsubsection{Differential Expression Analysis}
With the aim of comparing the results of the explainability phase (SHAP) with a canonical statistical approach, we performed a DEA on the complete datasets (i.e., before the declustering step). For microarray data we used the Wilcoxon rank-sum test adjusted by Benjamini–Hochberg, $FDR < 0.05$. For the scRNA-seq datasets we used the Seurat\cite{Seurat} implementation for DEA.  We then compared SHAP and DEA outputs, in terms of overlap both in general and for a restricted number of genes known to be involved in MS. These genes were extracted from the Multiple Sclerosis Gene Database (MSGD) \url{http://bio-bigdata.hrbmu.edu.cn/MSGD} \cite{10.1093/database/baae037}. We selected all genes from experiments including blood or CSF tissue (downloading the file "Regulation of RNA level"), resulting in 383 genes in total.

\subsubsection{Pathway analysis of important genes}
\label{sec:net}

SHAP-prioritized genes derived from the microarray analysis, CD4 CSF, and B-cell CSF were intersected to identify potential biomarkers detectable both in circulating blood cells and in cerebrospinal fluid. 

We computed the intersection between CD4 CSF and microarray genes, and between B-cell CSF and microarray. The union of these gene sets was subsequently analyzed using STRING to explore functional interaction networks. We removed non-connected genes and applied a graph-based clustering method, namely Markov Cluster Algorithm (MCL)\cite{van2000graph}. The algorithm groups together genes that are more densely connected to each other than to the rest of the network, thereby identifying putative functional modules. A key parameter of MCL is the inflation value, which controls the granularity of clustering (lower values produce fewer, larger clusters, whereas higher values generate more, smaller clusters).
To determine the optimal inflation parameter, we performed a perturbation-based stability analysis. Briefly, small random perturbations were introduced into the network, and clustering was repeated to evaluate how consistently genes were assigned to the same clusters. The stability of the resulting clusters was quantified, allowing us to select the inflation value that maximized robustness while preserving biological interpretability. 

For the genes in this network, we also studied closely their SHAP values, in the respective models. In particular, we were interested in the dependency between SHAP importance and gene expression values, for each gene. If SHAP importance grows as the gene becomes more expressed, then the gene is a possible risk factor for MS. Conversely, if the SHAP importance decreases and becomes negative as the gene expression level grows, then the gene could be a possible protective factor. We display the StringDB network with this information colour coded.

\section{Results}

\begin{figure}[h]
    \centering
    \includegraphics[width=1\textwidth]{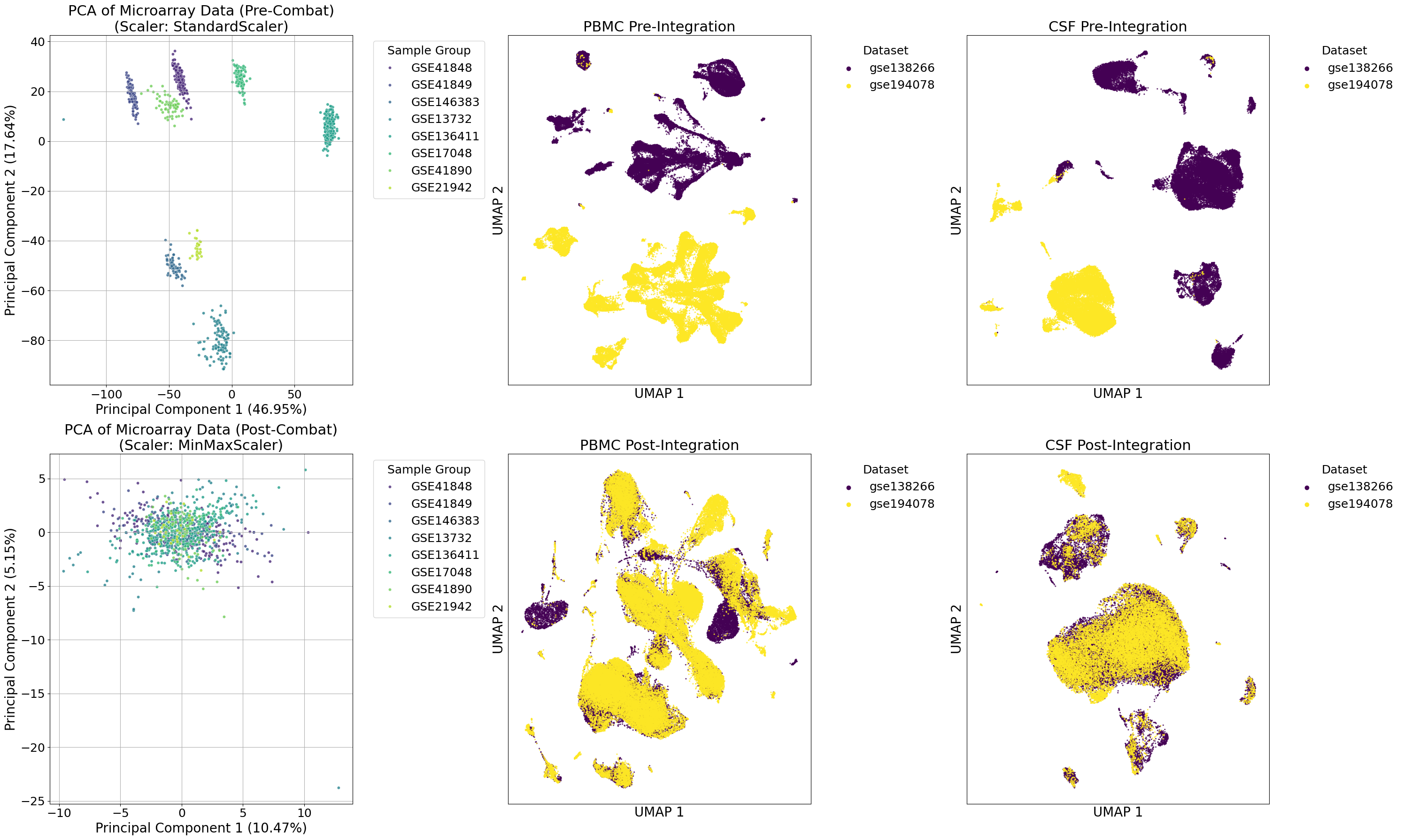}
    
    \caption{PCA/UMAP Pre- and Post-integration for the Microarray and the two scRNA-seq datasets. Different colours correspond to samples from different datasets.}
    \label{fig:PrePostIntegration}
\end{figure}

\subsection{Preprocessing}
\label{sec:preprocessingResults}
\paragraph{scRNA-seq}
The main summary statistics for the CSF and PBMC datasets are reported in Supplementary \autoref{table:TabellaDistribuzioneDatasetSingleCell}. Integration performance is shown in \autoref{fig:PrePostIntegration}, highlighting effective removal of batch effects and improved concordance across datasets, which enabled joint downstream analyses. From the integrated data, we extracted two immune cell subsets: CD4\textsuperscript{+} T cells and B cells.
Notably, the UMAP visualization indicates that dataset integration was more effective for CSF samples than for PBMC samples.

\paragraph{Microarray}
\label{chap:preprop_microarray}
After normalization and merging, the microarray dataset comprised $363$ control and $676$ MS samples (total, $1,038$ samples) and $6,673$ shared features. Inclusion of the GSE136411 and GSE17048 datasets substantially reduced the number of genes retained after harmonization ($6,673$ versus $11,492$), because these studies were generated on a different microarray platform (Illumina rather than Affymetrix), resulting in a smaller overlap of evaluated genes.

In \autoref{fig:PrePostIntegration}, we use principal component analysis (PCA) to illustrate how batch correction altered the global sample distribution relative to the uncorrected data. Visual inspection indicates that normalization markedly reduced dataset-specific separation. Consistent with this observation, differential expression analyses performed between individual datasets identified no statistically significant genes. In addition, the mean Mixture Score (averaged across the two conditions) was 0.29, where 0.50 represents the ideal value \cite{MixtureScore}.


\subsection{Declustering}
\paragraph{scRNA-seq}
In scRNA-seq data, the clustering algorithm identified a significant number of clusters of strongly correlated genes. 
The total number of clusters identified for each single-cell dataset and the average dimension of those clusters are available in \autoref{table:TabellaDistribuzioneClusterSingleCell} in the supplementary material.
When closely analyzing the clusters, we  observed how they consist of genes biologically related to each other. We chose the HLA-DRB1 gene as a benchmark and used it to identify the cluster in which said gene ended up being placed or becoming representative. This analysis showed us that it consistently formed clusters with other HLA-family genes.

\paragraph{Microarray}
In the microarray dataset, gene correlation analysis identified only one significant cluster, containing Y-linked genes: DDX3Y, EIF1AY, RPS4Y1, TXLNGY, UTY, PRKY, EIF1AY. The gene EIF1AY has been selected as representative (highest variance in the cluster), and the others were removed for model training.

\subsection{Training and Model Selection}
\label{ssec:trainingResults}



In \autoref{tab:XGBresults}, we can observe the performance of the best model identified for each dataset. In general, we observe a non-negligible difference between training and validation/test performance, indicating some overfitting affecting the models. However, test and validation performance remains good even if lower than training, and values are  quite similar between test and validation, indicating that our models are still able to generalise. To select only models with good overall performance, we decided to choose models with a test F1-score high enough to demonstrate good generalization capabilities on unseen data, while also displaying an AUC  $>0.60$, which strengthens our confidence that predictions are sufficiently better than random. Following these rules, we removed the model trained on CD4 PBMC, which did not demonstrate sufficient generalization capabilities, thus leading us to exclude it from further explainability and enrichment analysis. Among the remaining four models, the best performance was obtained on the B-cell CSF dataset, with test AUC$=0.94$ and F1$=0.9$, followed by the microarray (AUC$=0.86$ and F1$=0.75$) and the B-cell PBMC model (AUC$=0.73$, F1$=0.83$). The CD4 CSF model exhibited lower performance (AUC$=0.61$, F1$=0.71$) but still better than random so we preserved it. We note that on the scRNA-seq datasets, the CSF models behave better than the PBMC models, which could be in part due to poorer integration (see Figure~\ref{fig:PrePostIntegration}). Additionally, CD4 models perform worse than B-cell models, and we only maintain the PBMC one for subsequent analysis.  

\begin{table}[h]
\footnotesize
 \centering
    \caption{Performance of the top Gradient Boosting model per dataset.}
    \begin{tabular}{c||c||c||c|c|c|c|c}
    \textbf{Dataset} & \textbf{Train F1-Score} & \textbf{Validation F1-Score}&\textbf{Test F1-score} & \textbf{ Test Accuracy} & \textbf{Test Precision} & \textbf{Test Recall} & \textbf{Test AUC}\\
    \hline
     Microarray & 1.0 & 0.78 & 0.75 & 0.78 & 0.79 & 0.74 & 0.86\\
    \hline
     B cells CSF & 1.0 & 0.97 & 0.90 & 0.90 & 0.90 & 0.91 & 0.94\\
    \hline
     CD4 CSF & 0.99 & 0.78 & 0.71 & 0.69 & 0.66 & 0.62 & 0.61\\
    \hline
     B cells PBMC & 0.61 & 0.48 & 0.83 & 0.77 & 0.76 & 0.92 & 0.73\\
    \hline
     CD4 PBMC & 0.99 & 0.27 & 0.60 & 0.43 & 0.41 & 0.37 & 0.38\\
    \hline
    \end{tabular}
    \label{tab:XGBresults}
\end{table}

\begin{figure}[h!]
    \centering
    \thesubfigure {A. Best 15 features for SHAP importance in each dataset 
    \includegraphics[width=0.8\linewidth]{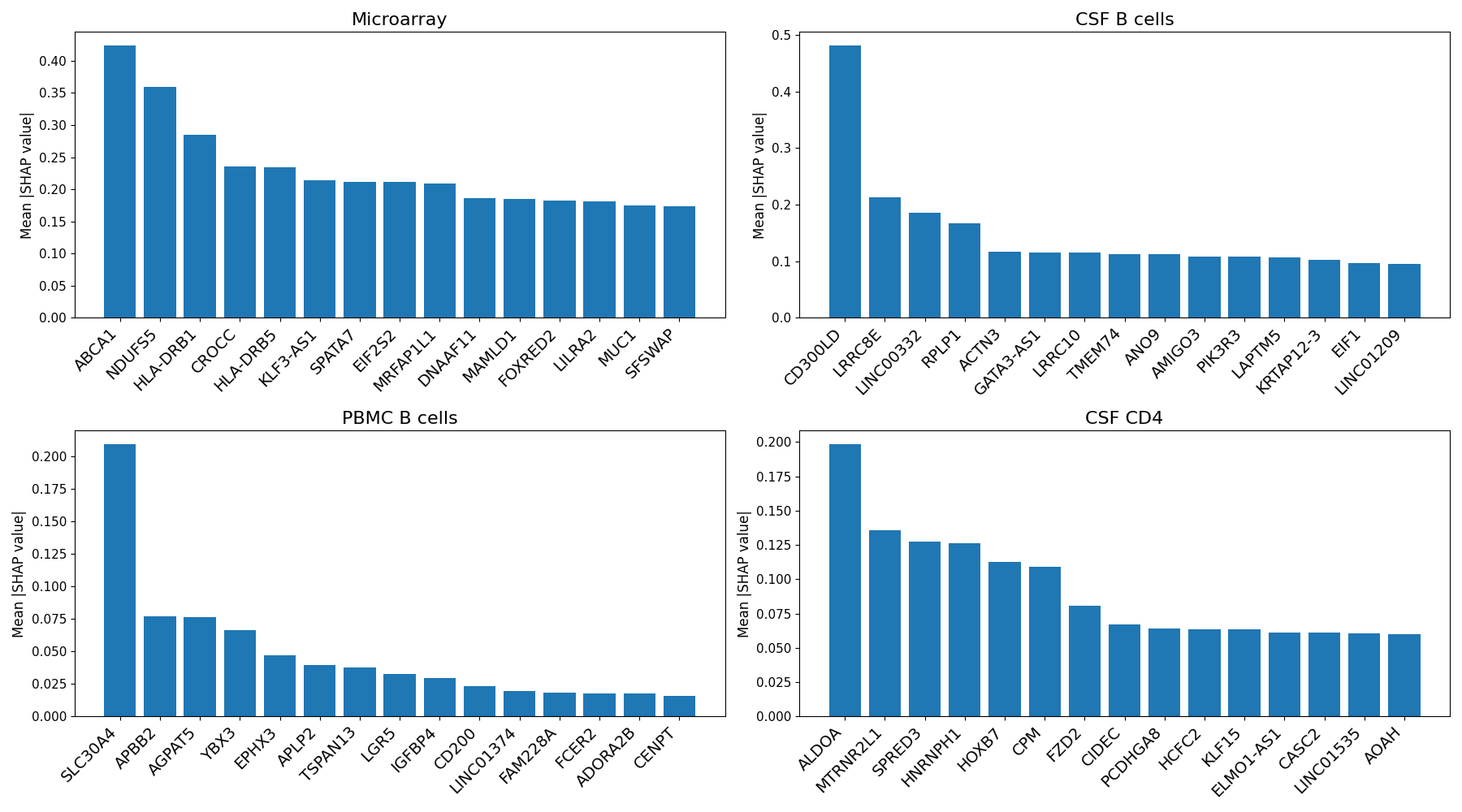}

   }
    \thesubfigure{ B. Venn diagram of genes identified by SHAP as important.
    
     \includegraphics[width=0.4\linewidth]{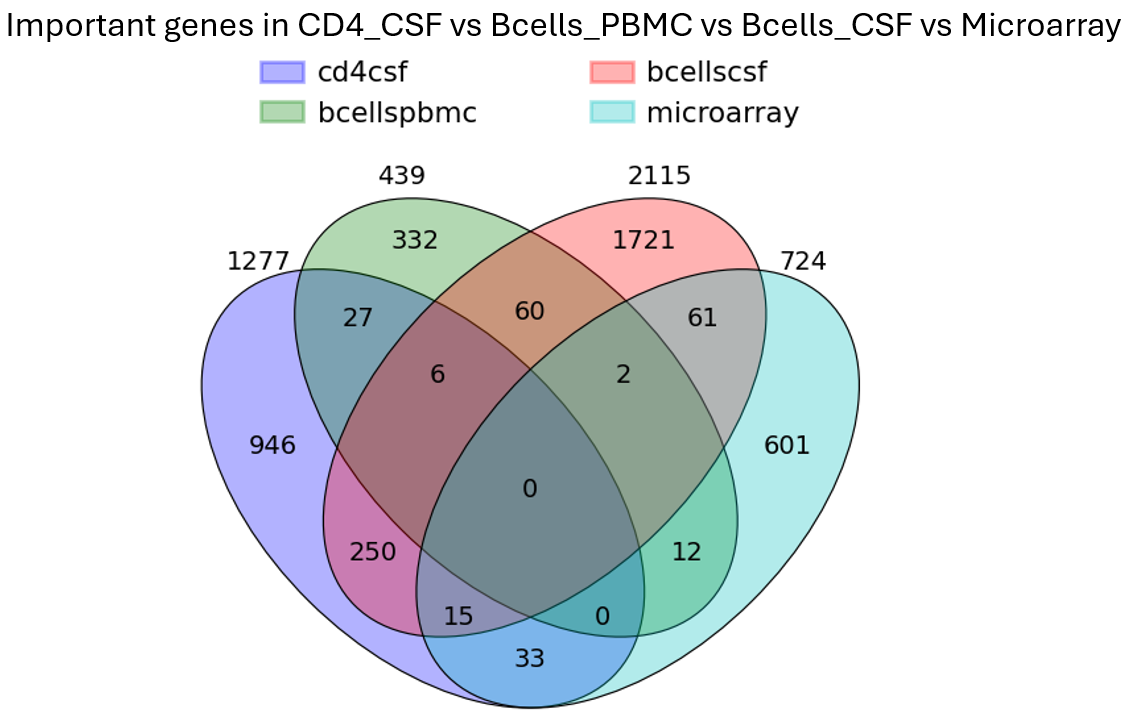}
    
    }
\caption{SHAP analysis: top 15 genes and the overlap among the four datasets.}
\label{fig:shap}
\end{figure}

\subsection{xAI \& Enrichment}

 \autoref{fig:shap}A displays the top 15 features for each dataset, ranked by the mean absolute importance from SHAP analysis. These restricted sets of genes do not show any overlap, indicating that biomarkers can be different across tissues and cell types. The size of the complete set of  genes selected for each dataset is presented in \autoref{tab:xAInumbers} in the supplementary material. We observe that this number varies significantly across the datasets; this phenomenon can be attributed to two main factors: the first is that the number and the size of gene clusters deeply depend on the dataset we are considering. Second, the feature importance distribution is inherently influenced by the model’s structure, particularly by the subset of features actively contributing to classification decisions. 
The Venn diagram in \autoref{fig:shap}B shows the intersection among the 4 gene sets. No gene is common to all 4 datasets. When considering only the scRNA-seq datasets, six shared genes emerge: AC104699.1, COL19A1, IGHA1, MACROD2, TSC22D1, UCHL1, where the latter is a known potential marker of the disease\cite{gorska2023uchl1}. 
PBMC datasets (1 microarray, 1 B-cell scRNA-seq) show only 14 genes in common, while overlap between microarray and both CSF datasets is much larger (15 genes in all 3 datasets, 33 in common between CD4 CSF and microarray, 63 between B-cell CSF and microarray).  When looking at pairs of datasets, the largest overlap is among the two CSF datasets (271 genes), indicating that tissue type is important in deciding biomarkers, even across cell types. 

\begin{figure}[h]
    \centering
    \includegraphics[width=\linewidth]{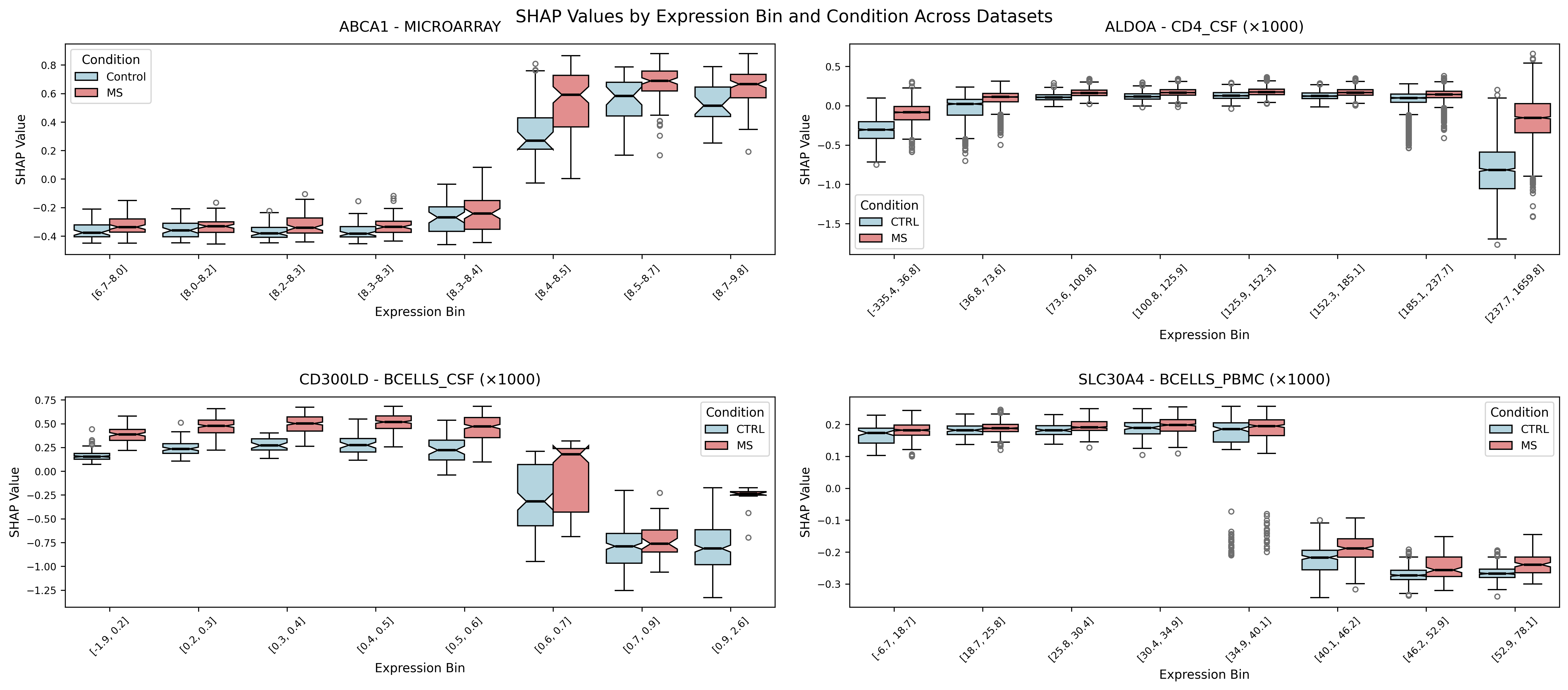}
    \caption{Dependence Analysis for Top Feature (based on SHAP Values) for each dataset. For scRNA-Seq datasets expression has been scaled by $10^3$ in order to be readable.}
    \label{fig:DependenceOrizzontaleTopFeatures1}
\end{figure}

\autoref{fig:DependenceOrizzontaleTopFeatures1} shows the distribution of importance values for the most important gene in each dataset, grouped by expression value and sample label (Control vs MS). Positive importance values indicate that in those samples, the gene increases the probability of MS, while negative values decrease it. For instance, in the microarray dataset, low ABCA1 expression levels appear to decrease the probability of MS, while high values increase it (in the top left panel of Figure~\ref{fig:DependenceOrizzontaleTopFeatures1} all boxplots on the left are below 0, while those on the right above 0), indicating ABCA1 as a risk factor. There appears to be a difference between Control (blue boxplots) and MS patients (red boxplots), with risk increased in MS patients.  

SLC30A4 exhibits opposite behaviour in the B-cell PBMC dataset, indicating a protective role, more marked in Control patients. This type of analysis, that concentrates gene importance in the model, based on gene expression levels, can be very useful to better understand the putative biomarker role in the disease, as we will show later in \autoref{sec:bio}. 

Enrichment analysis using the selected genes for each dataset showed some overlap across pathways. \autoref{tab:pathways} in supplementary material lists the top 10 pathways ranked by lowest FDR for each dataset, considering both KEGG and Reactome, while \autoref{fig:commonPathways} in the supplementary material displays the common significant pathways across the datasets. The number of significant KEGG pathways is lower than that of Reactome, although some biologically relevant pathways, such as Th1/Th2 and Th17 cell differentiation, were identified. The Reactome network reveals multiple pathways shared among Microarray, CSF\_B, and CSF\_CD4, while most dual overlaps occur between CSF\_B and CSF\_CD4. This observation is consistent with the fact that these cell populations were derived from the same substrate and measured with the same technology.

\subsection{DEA vs SHAP}
For the microarray dataset, the comparison between genes identified by SHAP and DEA showed an overlap of 167 genes. Besides those, 133 genes were uniquely identified by the SHAP analysis, while DEA  was much less specific and identified over 1,000 additional genes. Both methods identified known MS-associated genes such as HLA-DRB1 and HLA-DRB5, which ranked among the top five in SHAP importance. However, other genes with known associations, e.g. GABPA, MBP, and GFI1 were found only through DEA  (SHAP detects GFI1B, a strictly related gene), while SHAP uniquely highlighted EGR1, IL1B, and IL2RA. IL1B is a mediator of inflammation. Its blockade attenuated the disease in a pre-clinical model of MS\cite{linNewInsightsRole2017} while IL2RA is another well-known susceptibility gene, which was also used as a drug target in MS\cite{weberIL2RAIL7RAGenes2008, baranziniGeneticsMultipleSclerosis2017}. EGR1 is important in regulating the development of Treg and seems a key factor in neural survival in murine model of autoimmune encephalomyelitis\cite{yang2025early}.

For the scRNA-seq data, DEA applied to datasets  CD4 CSF, BCELLS PBMC, and BCELLS CSF resulted in 1018, 3925, and 727 genes, respectively.
Out of these, we had 145 genes in common with SHAP analysis in CD4 CSF, 287 in BCELLS PBMC, and 166 in BCELLS CSF, indicating significant coherence between the two methods. However, a large amount of extra information is provided by each method.

\begin{figure}[h]
    \centering
    \includegraphics[width=0.7\linewidth]{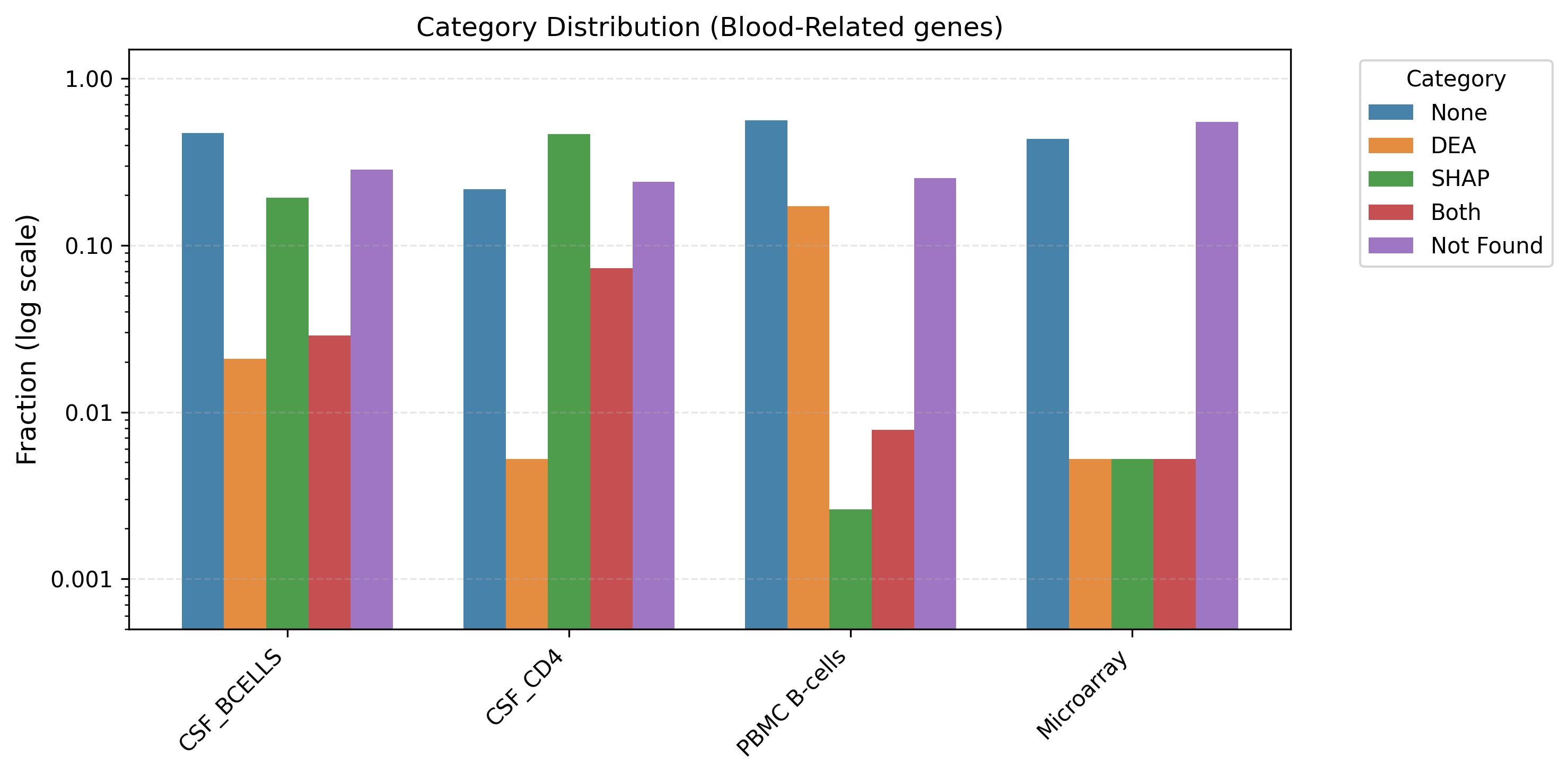}
    \caption{Fraction of genes known to be associated with MS that are identified as important by DEA and SHAP analysis. Some genes are not found in the datasets after cleaning and integration.}
    \label{fig:deavsshap}
\end{figure}

To further compare the two methods,  Figure~\ref{fig:deavsshap} shows the percentage of genes known to be associated with MS from the MSGD database (383 genes in total), identified by SHAP, DEA or both methods. We also include the actual list of genes with all results as a supplementary file. 

First of all, we note that a large number of genes are not included in our datasets (Not Found  category). Upon manual inspection, we observed that most of these genes are miRNAs. Therefore, they are rarely included or detected in mRNA experiments. In the microarray dataset, some genes are missing also due to integration: when combining 8 different datasets, we are left with only 6673 genes, as reported in Section~\ref{sec:preprocessingResults} above.

Concentrating on genes that exist in the datasets, for PBMC, we note that the two methods select a limited number of these genes, and they rarely agree. For the scRNA-seq, this can be also due to the poorer dataset integration (see Figure~\ref{fig:PrePostIntegration}) and the not so good model derived (see scores in Table~\ref{tab:XGBresults}) while for the microarray the large number of genes in the category "Not found" is due to the intrinsic small number of genes detectable by microarray. For microarray PBMC, the agreement is on genes HLA-DRB1 and HLA-DRB5, main biomarkers of MS, confirming the validity of our analysis. On B-cell PBMC, the agreement is on genes BACH2, HSP90B1 and FCRL1.
FCRL1 is a co-receptor of B-Cell Receptor (BCR) that modulate B-cell sensing and activation orchestrating humoral immune responses \cite{chen2025fcrl1}. HSP90B1 is an obligated chaperone protein for Toll-like-receptors (TLRs) folding. TLRs act as "secondary sensors" that work in tandem with BCR avoiding aberrant activations \cite{bernaleau2024ccdc134}.
The transcription factor BACH2 orchestrates both B-cell differentiation programs \cite{muto2010bach2} and the maintenance of lymphocyte stemness, thereby preventing terminal exhaustion of T cells and ensuring long-term cellular fitness \cite{chang2026bach2}.
On the same B-cells PBMC, DEA appears to identify more genes than SHAP, while SHAP identifies the gene SIL1. SIL1 (also known as BAP) acts as a nucleotide exchange factor for the chaperone BiP to ensure correct protein folding. Since IgG folding is resilient to BiP fluctuations, the SIL1-BiP axis likely supports a distinct set of proteins \cite{ichhaporia2015sil1}. Given their roles in BCR/TLR signaling integration, chaperone-mediated protein folding, and the control of lymphocyte differentiation and exhaustion, these genes may contribute to the aberrant immune activation and persistence of autoreactive responses that characterize Multiple Sclerosis.

For the two CSF datasets, the genes commonly detected by both SHAP and DEA are 11 in B-cells and 28 in CD4, while SHAP analysis seems to identify much more genes (74 genes in B-cells and 178 genes in CD4).  




 All in all, we observe that DEA and SHAP are complementary: some overlap exists and sometimes includes known biomarkers, while each methodology seems to add some extra knowledge. In particular, in our case SHAP identified more MS-related genes from single cell CSF data, while DEA was stronger on single cell PBMC data.

\subsection{Pathway analysis of SHAP prioritized genes}
\label{sec:bio}

\begin{figure}[t]
    \centering
    \includegraphics[width=1\textwidth]{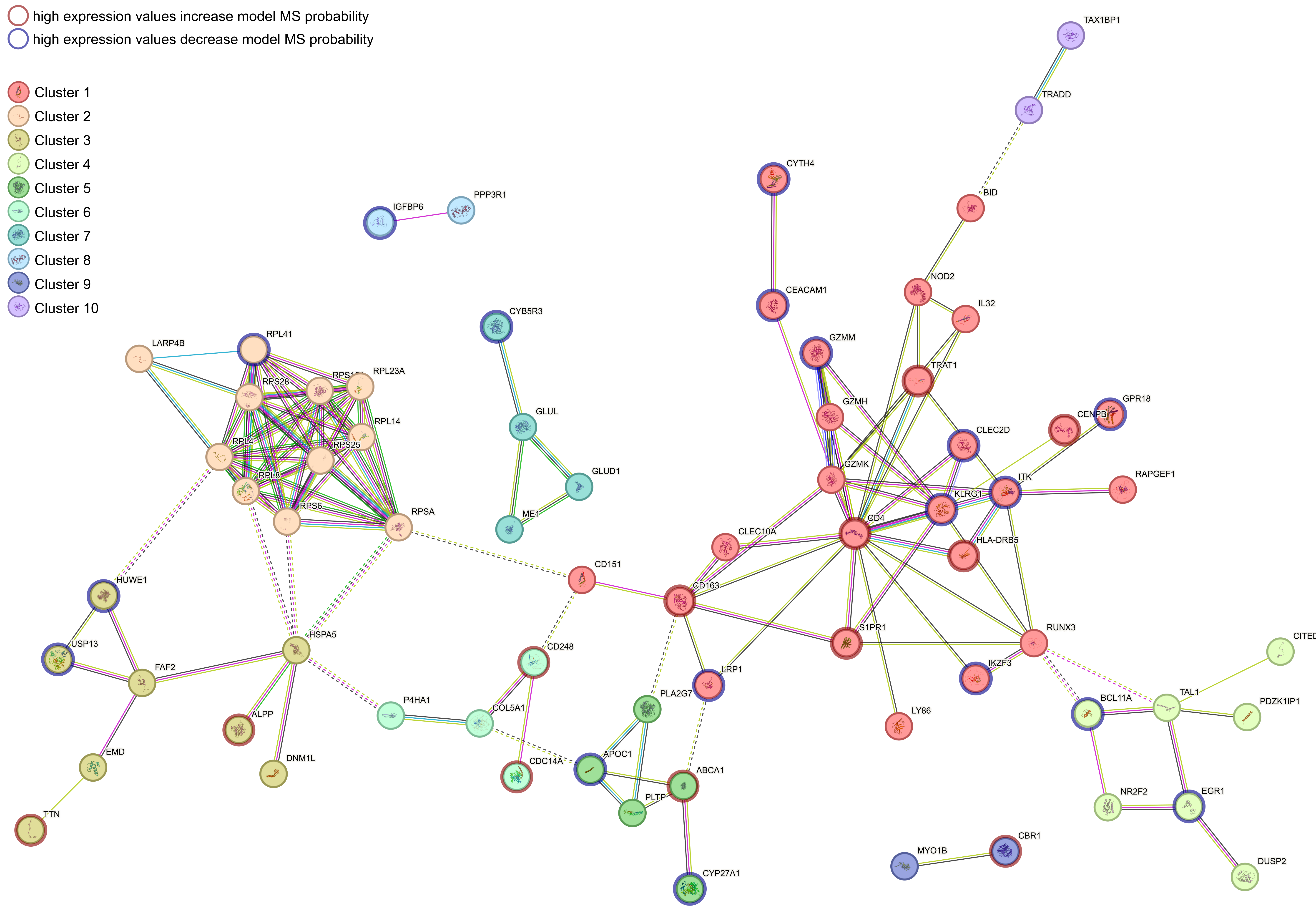}
    \caption{STRING network analysis. The ten clusters are identified by different node background colours. The node border colour, instead, indicates the effect of the genes on the model, as extracted from SHAP analysis. Intuitively, red borders indicate possible risk factors, genes for which a larger expression level increases MS probability, while blue borders indicate possible protective factors (see Methods subsection \ref{sec:net} for details).}
    \label{fig:stringNetwork}
\end{figure}

The overlap between CD4 CSF and the microarray dataset yielded 48 genes, whereas the overlap between B-cell CSF and the microarray dataset identified 78 genes. The union of these genes was studied in StringDB.
The perturbation-based stability analysis of clusters (see Methods subsection \ref{sec:net}) identified ten as the minimal number of clusters that showed both high stability across perturbations and clear biological relevance (see \autoref{fig:MCLcluster} in the supplementary material). We display the final network and the corresponding clusters in \autoref{fig:stringNetwork} and \autoref{tab:sdb}. 

Cluster 1 was characterized by an enrichment of genes involved in both adaptive and innate immune responses. The cluster included markers of T-cell activation (ITK, TRAT1, HLA-DRB5), differentiation (CD4, RUNX3, IKZF3) and trafficking (S1PR1, GPR18, CYTH4 and RAPGEF1, CD151) together with granzyme serine proteases and interleukins secreted by cytotoxic lymphocytes (GZMK, GZMH, GZMM, IL32), as well as key regulators of apoptosis such as BID ~\cite{white2025phosphorylation}.
In parallel, CLEC2D (also known as LLT1), KLRG1 and CEACAM1 molecules can act as immune checkpoints, while the expression of CD163, LY86 is associated with macrophage activation and tissue-remodeling processes. Overall, the composition of Cluster 1 delineates a coherent immune signature that integrates T-cell activation, differentiation, trafficking, interaction with innate immune cells activity and immune checkpoints. 

Cluster 2 was predominantly composed of genes encoding ribosomal proteins and factors associated with mRNA processing and translation. The cluster included multiple components of both the small (RPSA, RPS28, RPS25, RPS6, RPS15A) and large (RPL41, RPL23A, RPL14, RPL8, RPL4) ribosomal subunits, indicating a transcriptional program strongly enriched for protein synthesis and translational capacity. 
In addition, the presence of LARP4B, a regulator of mRNA stability and poly(A)-binding protein interactions, further supported the involvement of post-transcriptional control mechanisms. Overall, the composition of this cluster delineates a ribosomal/translation signature.

Cluster 3 comprises genes involved in protein homeostasis, mitochondrial dynamics, and cellular structural integrity. Key components of the Endoplasmic Reticulum (ER) stress and proteostasis machinery (HSPA5, HUWE1, USP13, FAF2) suggested activation of protein quality-control pathways, while DNM1L indicated an involvement of mitochondrial fission. 
EMD and TTN reflected changes in nuclear and cytoskeletal architecture. Overall, Cluster 3 represents a transcriptional program linking ER stress responses, ubiquitin-mediated proteostasis, mitochondrial dynamics, and structural organization.

Cluster 4 was composed of transcriptional regulators and early-response genes involved in cell fate, differentiation, and signaling modulation. Key factors such as CITED2, TAL1, BCL11A, and NR2F2 indicated a strong contribution of developmental and lineage-specification programs, while EGR1 and DUSP2 reflected rapid transcriptional responses to external stimuli and regulation of MAPK signaling. The presence of PDZK1IP1 in this cluster, associated with growth and stress-adaptive pathways, further supported a role in dynamic transcriptional remodeling. Overall, this cluster highlights a coordinated signature of transcriptional control, stimulus responsiveness, and regulation of cell signaling.

Cluster 5 contains genes involved in extracellular matrix organization and lipid metabolism. COL5A1 pointed to active matrix remodeling, while APOC1, PLTP, ABCA1, CYP27A1, and PLA2G7 indicated coordinated regulation of lipid handling, including lipoprotein remodeling, cholesterol efflux, and phospholipid metabolism. 
Overall, this cluster reflects a transcriptional program linking matrix organization with lipid transport and inflammatory lipid-processing pathways.

Cluster 6 was enriched for genes linked to extracellular matrix organization and stromal remodeling. P4HA1 and COL5A1 indicate active collagen synthesis and structural matrix modulation, while CD248 supports mesenchymal and perivascular cell activation associated with tissue remodeling. The presence of CDC14A suggests concurrent regulation of cell-cycle dynamics and cytoskeletal organization. Overall, this cluster reflects a compact transcriptional program related to extracellular matrix remodeling and structural adaptation within an inflammatory context.

Cluster 7 included  genes involved in cellular metabolism, particularly amino-acid and redox pathways. GLUD1 and GLUL pointed to active glutamine–glutamate cycling and nitrogen handling, while CYB5R3 and ME1 reflected enhanced redox balance and NADPH generation. Recent Metabolomic analysis highlighted mitochondrial dysfunction in the enriched pathways after adjustment for MS status, with CYB5R3 identified as a key gene in the metabolism of phosphatidylcholine and potentially contributing to altered mitochondrial activity~\cite{andersen2019metabolome}.
Overall, this cluster highlights a metabolic signature centered on glutamine metabolism, redox regulation, and cellular energy support.

Cluster 8 is a small cluster that includes only two genes involved in growth factor modulation and calcium-dependent signaling. IGFBP6 indicated regulation of insulin-like growth factor availability, while PPP3R1, the regulatory subunit of calcineurin, pointed to modulation of calcium-dependent signal transduction. 

Similarly, cluster 9 was composed of only two genes linked to cytoskeletal dynamics and cellular redox metabolism. MYO1B reflected actin-based membrane remodeling and vesicular transport, while CBR1 indicated involvement in oxidative stress regulation and NADPH-dependent carbonyl metabolism. Finally cluster 10 is involved in apoptotic regulation and TNF-mediated inflammatory signaling and NF-kB–related signaling cascades. TRADD is a key adaptor in TNF receptor pathways, highlights activation of death-receptor  while TAX1BP1 points to modulation of NF-kB activity and selective autophagy processes that regulate immune responses and cellular stress. Overall, this cluster contains genes centered on inflammatory signaling control and apoptosis-related pathways. 

Having defined the ten clusters and their functional signatures, we next contextualize these findings within the current knowledge of MS pathogenesis.

\begin{table}[h!]
\centering
\footnotesize
\caption{Description of clusters obtained by STRING analysis on SHAP prioritized genes. *indicate manual curated pathway description (not directly obtained by STRING)}
\label{tab:sdb}

\begin{tabular}{|p{1cm}|p{1cm}|p{0.6cm}|p{0.27\textwidth}|p{0.48\textwidth}|}
\hline
\textbf{Cluster number} & \textbf{Cluster color} & \textbf{Gene count} & \textbf{Description} & \textbf{Protein names} \\
\hline

1 & Red & 25 & *T-cell activation, differentiation, trafficking, interaction with innate immune cells activity and immune check-points &
CD4, LY86, GZMM, IL32, LRP1, CLEC2D, CEACAM1, IKZF3, CLEC10A, GZMH, RUNX3, GZMK, S1PR1, NOD2, KLRG1, CD163, ITK, TRAT1, HLA-DRB5, CYTH4, CENPB, BID, CD151, GPR18, RAPGEF1 \\
\hline

2 & Brown & 11 & Cytoplasmic translation, Viral mRNA Translation, Ribosome &
RPL8, LARP4B, RPL41, RPS15A, RPL4, RPS28, RPS6, RPS25, RPL23A, RPSA, RPL14 \\
\hline

3 & Olive & 8 & *ER stress, ubiquitin–proteasome regulation, and structural organization &
FAF2, EMD, USP13, HUWE1, HSPA5, DNM1L, ALPP, TTN \\
\hline

4 & Lime green & 7 & *Transcriptional control and signaling regulation &
CITED2, TAL1, BCL11A, NR2F2, EGR1, DUSP2, PDZK1IP1 \\
\hline

5 & Green & 5 & Cholesterol metabolism, Plasma lipoprotein particle organization &
ABCA1, CYP27A1, PLA2G7, PLTP, APOC1 \\
\hline

6 & Mint green & 4 & *Extracellular matrix remodeling &
P4HA1, COL5A1, CD248, CDC14A \\
\hline

7 & Teal green & 4 & Nitrogen metabolism, Glutamate catabolism & GLUD1, ME1, GLUL, CYB5R3 \\
\hline

8 & Light sky blue & 2 & *Growth factor modulation and calcium-dependent signaling & IGFBP6, PPP3R1 \\
\hline

9 & Blue & 2 & *Actin-based membrane remodeling, vesicular transport \& oxidative stress regulation & MYO1B, CBR1 \\
\hline

10 & Lavender Purple & 2 & *Inflammatory signaling control and apoptosis-related pathways & TRADD, TAX1BP1 \\
\hline

\end{tabular}

\label{tab:clusterSummary}
\end{table}

\section{Discussion}


Our integrative analysis identified a network composed of 10 gene clusters spanning distinct biological functions and pathways. Several genes within these clusters have previously been implicated in MS, supporting the biological relevance of the identified modules.
Cluster 1 was primarily associated with T-cell activation and effector functions. Notably, this cluster included S1PR1, a well-established therapeutic target of sphingosine-1-phosphate receptor modulators such as fingolimod and siponimod, which reduce lymphocyte trafficking by inducing receptor internalization~\cite{kandjani2023s1pr1}. 
The same cluster also contained CD163, a marker of macrophage activation, whose soluble form has been indicated as a potential biomarker in serum and cerebrospinal fluid of patients with MS ~\cite{stilund2014soluble, stilund2015biomarkers}.

An interesting module of genes within this cluster is composed by ITK, CLEC2D, KLRG1 and CEACAM1. 
ITK is a modulator of the strength of T cell receptor signaling. It has been shown that ITK inhibition can enhance CAR-T cell persistence, reducing T cell exhaustion ~\cite{fu20252}. Thus, targeted inhibition or deletion of ITK attenuates excessive activation signals, thereby preventing terminal differentiation and the onset of transcriptional exhaustion. This shift promotes the enrichment of memory subsets instead.
CLEC2D encodes for LLT1, and is the endogenous inhibitory ligand of CD161. Increased LLT1 expression has been found in inflammatory cells in chronic autoimmune diseases ~\cite{chalan2015expression}. The LLT1-CD161 axis is involved in the homing of T lymphocytes to immune-privileged sites, a critical step for CD161+ cells in MS pathogenesis. Interest in this pathway has also grown in cancer immunity, as a non-canonical immune checkpoint ~\cite{alvarez2024targeting}.
Also KLRG1 has been proposed as an immune checkpoint acting as a molecular brake inhibiting effector functions of immune cells. Overexpression of this gene in T cells during terminal differentiation and senescence leads to immune dysfunction. Studies suggest that KLRG1 is mostly positively correlated with disease severity in autoimmune diseases~\cite{zhang2024role}.
Within non-canonical immune check points we also found CEACAM1. Its interaction with TIM-3 drives immune tolerance and T cells to exhaustion ~\cite{huang2015ceacam1}. At the same time, this critical checkpoint for T cells may also facilitate the formation of pathogenic B-cell aggregates within the CNS ~\cite{rovituso2016ceacam1}. This CEACAM1 bifunctionality explains its diverse roles: it acts as a molecular chaperone for TIM-3, stabilizing the inhibitory complex required for T-cell exhaustion, while simultaneously functioning as a homophilic adhesion molecule that drives pathogenic B-cell aggregation in the CNS of MS patients.
The expansion and aggregation of CEACAM1+ B cells in MS could represent a failed compensatory attempt to induce anergy in autoreactive T cells via trans-engagement with TIM-3. This mechanism, intended to bypass impaired T-cell checkpoint signaling, ultimately fails and instead triggers pathological B-cell aggregation within the CNS. B-cell clusters become a maladaptive byproduct that accelerates disease progression~\cite{rovituso2016ceacam1}. 
Overall, this evidence suggested that CEACAM1 act as immune checkpoint molecules on autoreactive B and T cells identifying CEACAM1 as a clinically highly interesting target in MS pathogenesis central nervous system autoimmunity.
Interestingly, our model suggests that elevated levels of ITK, KLRG1, CLEC2D, and CEACAM1 might decrease the probability of MS. Consequently, the upregulation of this module might act as a limiting factor against the establishment of chronic neuroinflammation.

Besides the immune-related module, additional clusters highlighted transcriptional regulation.
Interestingly, in cluster 4, we identified EGR1 as a protective gene in multiple sclerosis, consistent with its role as a crucial regulator of regulatory T cells~\cite{yang2025early}. Moreover, CITED2 and BCL11A were identified as susceptibility genes by genome-wide association studies in MS ~\cite{baranzini2009genome}. 
EGR1 acts as a critical driver for Foxp3+ Treg differentiation, effectively suppressing MS neuroinflammation, while BCL11A is essential for lymphopoiesis differentiation of B cells and survival \cite{yu2012bcl11a}. 
These regulatory networks may contribute to immune cell homeostasis to prevent autoimmune chronic neuroinflammation.

Interestingly, cluster 2 was enriched for many ribosomal proteins and RPS6 has already been identified as a biomarker in multiple sclerosis~\cite{parnell2014ribosomal}. Targeting ribosomal proteins is emerging as an approach for modulating immune alterations in brain diseases~\cite{qi2023single}. Moreover, EBV Nuclear Antigen 1 (EBNA1), which is essential for EBV persistence, interacts with ribosomal protein L4 (RPL4) and nucleolin to stabilize viral genome maintenance, potentially contributing to chronic immune activation in the disease~\cite{shen2016ribosome}. Nowadays EBV is the most evident risk factor for MS as shown by a recent study that associate EBV infection with an increased risk of developing MS ~\cite{bjornevik2022longitudinal}.

Cluster 3 highlights ER stress and post-translational modifications. 
Increased levels of GRP78, which encodes the protein BiP, a chaperone that ensures correct protein folding in B cells \cite{ichhaporia2015sil1}, were found to influence the risk of MS \cite{maghbooli2025interplay}. Moreover, the interplay between the E3 ubiquitin ligase HUWE1 and the deubiquitinase USP13 emerges as a critical regulatory axis for protein stability within immune cells. While HUWE1 facilitates the proteasomal degradation of key inflammatory mediators, USP13 counteracts this process by removing ubiquitin chains, thereby preventing their destruction. The imbalance of the HUWE1-USP13 axis can lead to the pathological stabilization of pro-inflammatory factors, driving the chronic neuroinflammation characteristic of MS. Dysregulation of deubiquitinating enzymes in multiple sclerosis may sustain aberrant immune activation and contribute to chronic neuroinflammation ~\cite{li2025usp13}, highlighting the therapeutic potential of modulating the ubiquitin–proteasome system in autoimmune diseases~\cite{yadav2022modulating}.

Another interesting axis is highlighted by cluster 5. ABCA1 and APOC1 are  mediators of lipid trafficking and cholesterol homeostasis and the ABCA1/ApoA-I axis exerts potent anti-inflammatory effects by modulating lipid raft composition and hindering pathogenic interactions between T cells and macrophages. This mechanism facilitates the clearance of neurotoxic hydroxysterols across the blood-brain barrier, highlighting the importance of reverse cholesterol transport in maintaining immune privilege within the neuroinflammatory landscape~\cite{gardner2015importance}. It is also known that vitamin D and omega-3 fatty acids can influence these processes by modulating lipid-related pathways involving APOA1 and ABCA1~\cite{abd2025role}.  
In the same cluster there are involved extracellular matrix (ECM) proteins, including COL chains, that can modulate immune and glial cell functions~\cite{mohan2010extracellular}. 

This analysis was based on genes identified as important by a set of machine learning models. It is important to mention that the prediction performance of these models varied from one data type to another (as shown earlier in Table~\ref{tab:XGBresults}). In some cases performance was extremely good (B-cell CSF), while in others we observed some overfitting (stronger in CD4 CSF, lower in Microarray). This is most probably due to reduced data size, typical to the medical domain, and could be improved with additional data. However,  we underline the fact that in all cases test AUC was high enough, significantly above random. 

All in all, our findings delineate the multifaceted molecular landscape where MS pathogenesis is driven by a dysregulation of immune and metabolic checkpoints. The identification of clusters ranging from non-canonical immune checkpoints to lipid trafficking, protein post-translational modifications, and EBV infection underscores the complexity of neuroinflammation. Our model suggests that MS arises from a systemic failure to maintain immune tolerance and metabolic privilege within the CNS. These interconnected pathways offer a promising frontier for biomarker development and therapeutic interventions.

\section{Conclusion}
In this study we developed a machine learning pipeline to analyze MS microarray and scRNA-seq, including two tissue types (PBMC and CSF) and two immune cell types for scRNA-seq data (CD4 and B-cells). We trained models to classify MS and control patients in each dataset. Explainability analysis through SHAP allowed us to identify important genes in each data type and tissue. We compared selected genes with canonical DEA, observing some overlap and wide complementarity among the two methods.  SHAP generated more plausible hits in CSF data, while DEA was more successful in PBMC data (which could also be due to issues with integration of PBMC data).  Enrichment analysis of SHAP-detected genes identified plausible pathways in each dataset. Further network cluster analysis of overlapping genes underlined clusters ranging from non-canonical immune checkpoints to lipid trafficking, protein post-translational modifications, and EBV infection. The findings underscore the complexity of neuroinflammation of MS and describe how MS arises from a systemic failure to maintain immune tolerance. 

The four immune checkpoints identified by our analysis (ITK, CLEC2D, KLRG1 and CEACAM1) can be used as hypotheses for new biomarkers or targets for therapy. In future work we plan to include these in mechanistic models of MS~\cite{10.1093/bioinformatics/btaf103}, to validate this hypothesis. We will also concentrate on these genes or the pathways in which they are involved in further multiomics studies based on new MS samples. On the data analysis methodological side, future work will include the exploration of existing foundation models for transcriptomics data and their application to MS, and the development of MS-specific foundation models.

\clearpage
\section*{\bf Conflict of interests}
\label{sec:CONFLICT-OF-INTERESTS}
The authors declare no conflict of interest.

\section*{Author contributions statement}

S.G.G., S.R., M.B., M.C., C.P. and A.S. conceived the experiment, all authors designed the analysis pipeline, F.M., S.P.(1), S.G.G. and A.S. selected the datasets, F.M. and S.P.(1) implemented the preprocessing and ML  pipeline and applied it to the datasets. A.M. and S.P.(2) performed network analysis and clustering. S.G.G. and A.M. performed the biological analysis of results. All authors interpreted the results and contributed to drafting the manuscript. All authors approved the final manuscript.

\section*{Additional information}

\section*{\bf Acknowledgments}
We gratefully acknowledge the support of the CINI InfoLife laboratory in this research.

\section*{\bf Funding}
\label{sec:FUNDING}
This work was supported by the co-funding European Union - Next Generation EU, in the context of The National Recovery and Resilience Plan, Mission 4 Component 2, Investment 1.1, Call PRIN 2022 D.D. 104 02-02-2022 – MEDICA Project, CUP N. I53D23003720006 \& D53D23008800006. We also acknowledge partial support from Investment 1.5 Ecosystems of Innovation, Project Tuscany Health Ecosystem (THE), CUP: B83C22003920001, Spoke 3, from the SPARK programme at the University of Pisa, from the project "Hub multidisciplinare e interregionale di ricerca e sperimentazione clinica per il contrasto alle pandemie e all’antibioticoresistenza (PAN-HUB)” funded by the Italian Ministry of Health (POS 2014-2020, project ID: T4-AN-07, CUP: I53C22001300001) and from a 2023 NARSAD Young Investigator Grant from the Brain \& Behavior Research Foundation.

\section*{\bf Availability of data and software code}
\label{sec:AVAILABILITY}
Code and data available at: 
 \indent \url{https://github.com/seriph78/ML_for_MS.git}
\\\\ 
The datasets analyzed during this study were obtained from the GEO repository, as follows: 
\\ 
Microarray data:
 GSE41848: \hyperlink{https://www.ncbi.nlm.nih.gov/geo/query/acc.cgi?acc=GSE41848}{link}, 
 GSE41849: \hyperlink{https://www.ncbi.nlm.nih.gov/geo/query/acc.cgi?acc=GS   E41849}{link}, 
GSe146383: \hyperlink{https://www.ncbi.nlm.nih.gov/geo/query/acc.cgi?acc=GSE146383}{link}, 
 GSE13732: \hyperlink{https://www.ncbi.nlm.nih.gov/geo/query/acc.cgi?acc=GSE13732}{link}, 
 GSE136411: \hyperlink{https://www.ncbi.nlm.nih.gov/geo/query/acc.cgi?acc=GSE136411}{link}, 
 GSE17048: \hyperlink{https://www.ncbi.nlm.nih.gov/geo/query/acc.cgi?acc=GSE17048}{link}, 
GSE41890: \hyperlink{https://www.ncbi.nlm.nih.gov/geo/query/acc.cgi?acc=GSE41890}{link}, 
 GSE21942: \hyperlink{https://www.ncbi.nlm.nih.gov/geo/query/acc.cgi?acc=GSE21942}{link}.
\\
scRNA-seq data:
GSE138266: \hyperlink{https://www.ncbi.nlm.nih.gov/geo/query/acc.cgi?acc=GSE138266}{link}, 
GSE194078: \hyperlink{https://www.ncbi.nlm.nih.gov/geo/query/acc.cgi?acc=GSE194078}{link}.

\newpage
\bibliography{Bibliography.bib} 
\normalsize

\newpage
\setcounter{figure}{0}
\setcounter{table}{0}
\section*{\huge Supplementary Material}

\vspace{2em}
\hrule
\vspace{1em}

\begin{figure}[b]
    \centering
    \includegraphics[width=\linewidth]{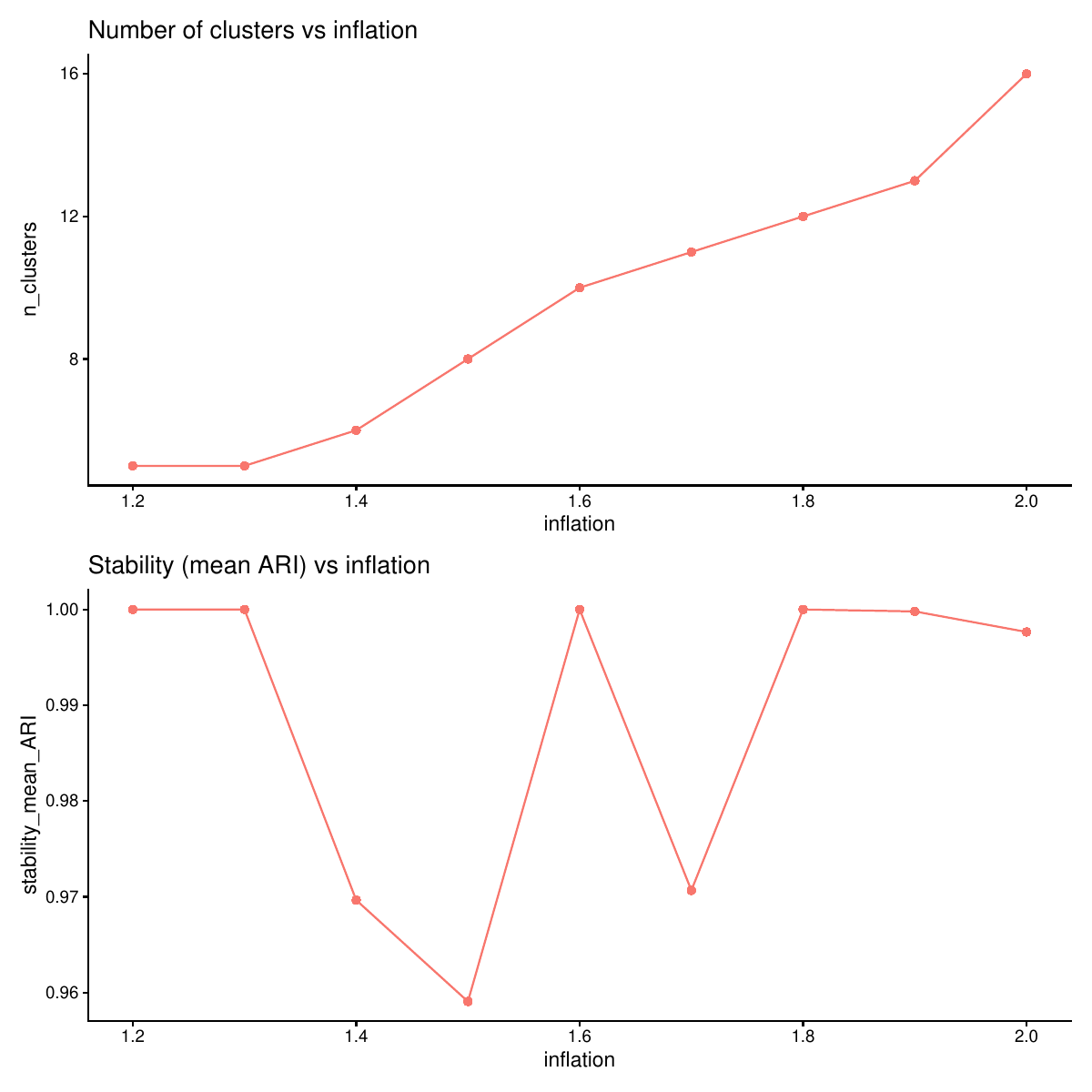}
    \caption{MCL clustering analysis}
    \label{fig:MCLcluster}
\end{figure}

\begin{figure}[t]
    \centering
    \subfloat[Common KEGG pathways across datasets]{
        \includegraphics[width=0.44\linewidth]{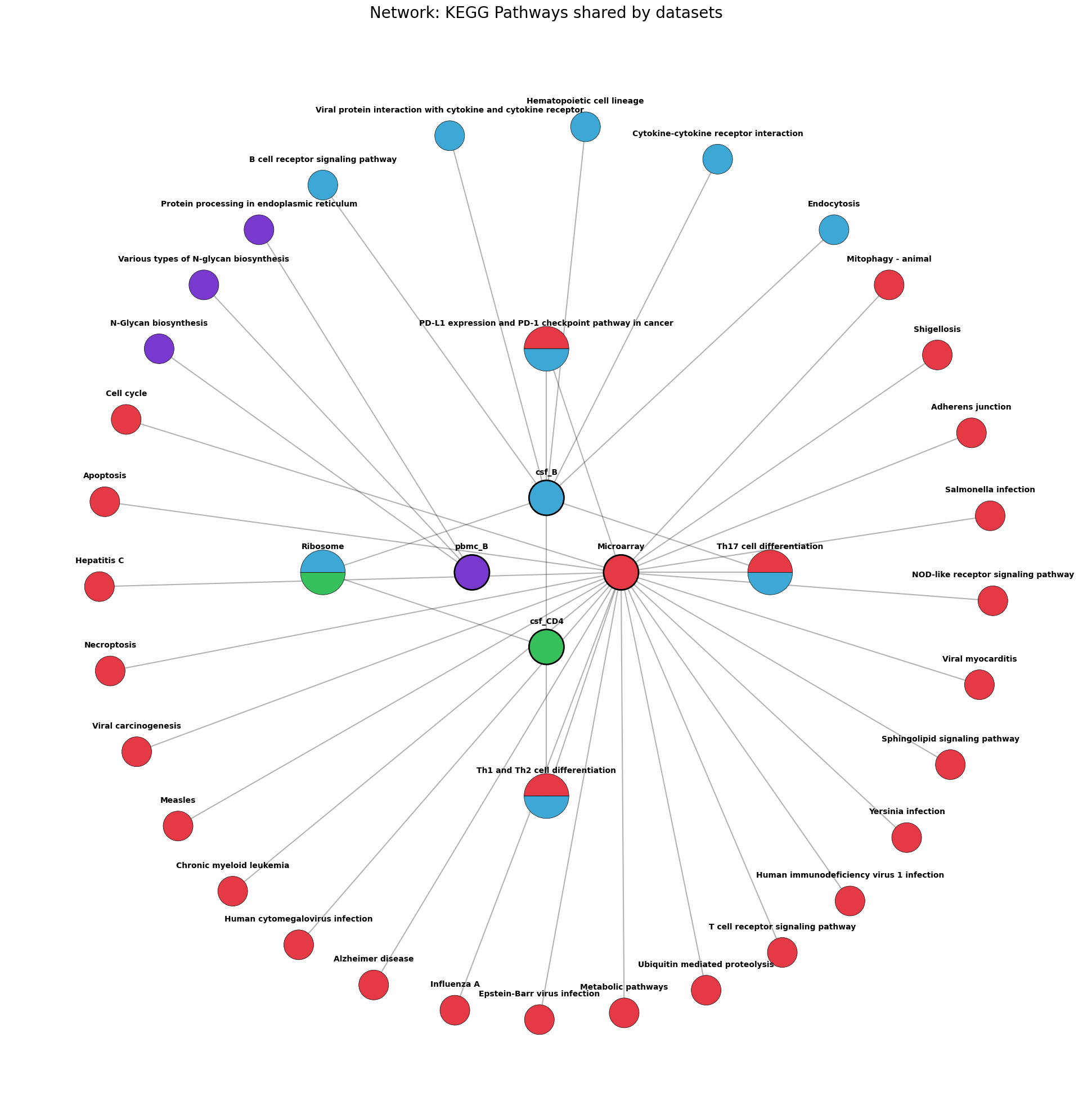}
        \label{fig:keggPathways}
    }
    \hfill
    \subfloat[Common Reactome pathways across datasets]{
        \includegraphics[width=0.54\textwidth]{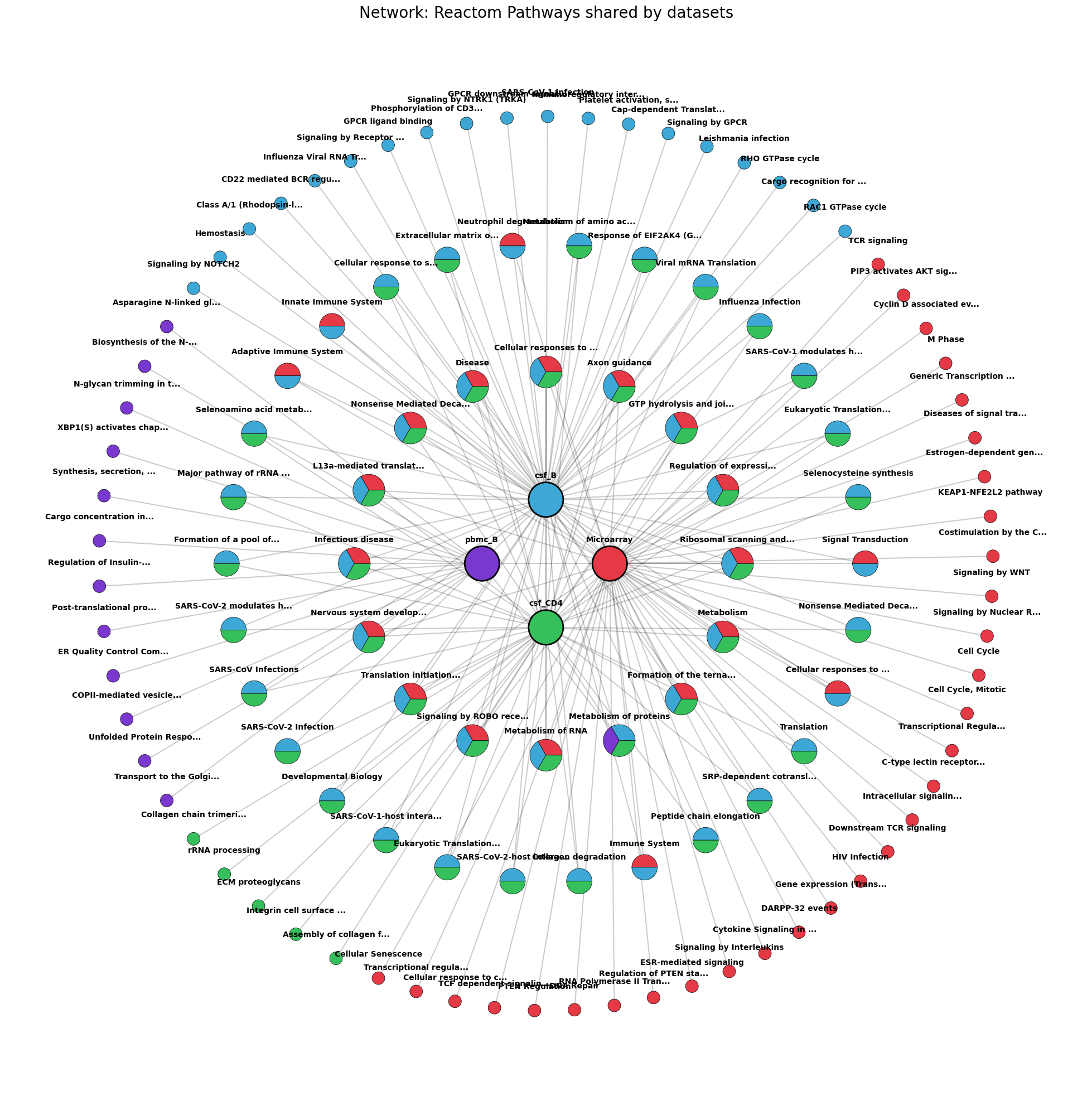}
        \label{fig:reactomePathways}
    }
    \caption{Common pathways across datasets}
    \label{fig:commonPathways}
\end{figure}

\begin{figure}
    \centering
    \includegraphics[width=1\linewidth]{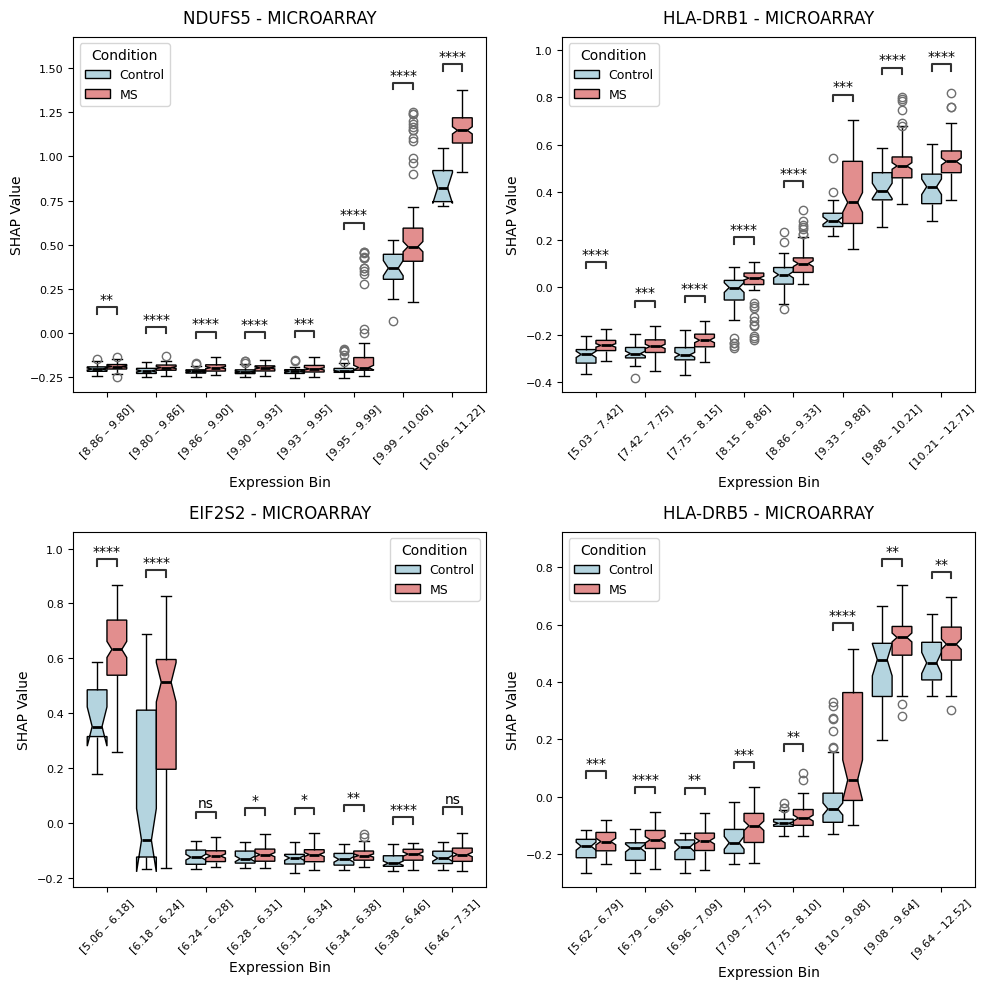}
    \caption{Box plot and tests considering some interesting genes for microarray}
    \label{fig:dp}
\end{figure}

\textbf{\Large Contents}
\\
\begin{itemize}
    \large{
    \item Single Cell train, validation and test distributions- Table~\ref{tab:divisione_training_val_test_singlecell_fixed}
    \item scRNA-seq and Microarray hyperparameter search space  - Table~\ref{tab:scRNAseqHyperparameters} and Table~\ref{tab:microarrayHyperparameters}
    \item Single Cell datasets statistics - Table~\ref{table:TabellaDistribuzioneDatasetSingleCell}
    \item Result of cluster analysis - Table~\ref{table:TabellaDistribuzioneClusterSingleCell}
    \item Models hyperparameters choosen via Bayesian Optimization - Table~\ref{tab:ChoosenHyperparameters}
    \item Number of important feature retrieved after SHAP for each dataset - Table~\ref{tab:xAInumbers}
    \item Pathway enrichment for genes selected by SHAP analysis in each dataset - top 10 pathways - Table~\ref{tab:pathways}
\item Cluster analysis of interaction graph - Figure~\ref{fig:MCLcluster}
\item Overlap of enriched pathways across datasets. - Figure~\ref{fig:commonPathways}
\item Dependency plot for some genes in microarray data - Figure~\ref{fig:dp}
    \item SHAP and DEA results for the MSGD genes.
    }
\end{itemize}

\begin{table}[h]
    \footnotesize
    \centering
    \caption{Distribution of Cells Across Training, Validation, and Test Sets for Single-cell Datasets}
    \begin{tabularx}{\textwidth}{|c|X|X|X|X|X|X|}
        \hline
        \textbf{Dataset} & \textbf{Train\newline MS Cells} & \textbf{Train\newline Control Cells}
         & \textbf{Validation\newline MS Cells}
         & \textbf{Validation\newline Control Cells}
          & \textbf{Test\newline MS Cells}
           & \textbf{Test\newline Control Cells} \\
        \hline
        CD4 PBMC & 7728 & 4019 & 1499 & 1012 & 1584 & 2074  \\
        CD4 CSF & 11200 & 5263 & 3523 & 1106 & 3129 & 1463  \\
        BCELLS PBMC & 1077 & 439 & 168 & 398 & 876 & 547  \\
        BCELLS CSF & 809 & 583 & 171 & 177 & 130 & 228  \\
        \hline
    \end{tabularx}
    \label{tab:divisione_training_val_test_singlecell_fixed}
\end{table}

\begin{table}[h]
    \centering
    \begin{minipage}[t]{\textwidth}  
    \footnotesize
    \caption{scRNA-seq hyperparameter search space.}
    \centering
    \begin{tabular}{c|c||c|c}
        \textbf{Hyperparameter} & \textbf{Range}     & \textbf{Hyperparameter}     & \textbf{Range}             \\
        \hline
        \textit{n\_estimators}  & $50$ to $3000$     & \textit{min\_child\_weight} & $1$ to $8$                 \\
        \hline
        \textit{max\_depth}     & $2$ to $15$        & \textit{scale\_pos\_weight} & $\{1, \#min\_class \; / \; \#maj\_class \}$\\
        \hline
        \textit{learning\_rate} & $0.0000001$ to $1$ & \textit{reg\_alpha}         & $0.000001$ to $100$        \\
        \hline
        \textit{gamma}          & $0.000001$ to $1$  & \textit{reg\_lambda}        & $0.000001$ to $100$        \\
        \hline
    \end{tabular}
    \label{tab:scRNAseqHyperparameters}
    \end{minipage}
    
    \hfill
    
    \begin{minipage}[t]{\textwidth}  
    \footnotesize
    \caption{Microarray hyperparameter search space.}
    \centering
    \begin{tabular}{c|c||c|c}
        \textbf{Hyperparameter} & \textbf{Range} &  \textbf{Hyperparameter} & \textbf{Range} \\
        \hline
        \textit{n\_estimators} & $50$ to $600$ &    \textit{reg\_alpha} & $0.0001$ to $100$ \\
        \hline
        \textit{max\_depth} & $2$ to $15$ & \textit{reg\_lambda} & $0.0001$ to $100$\\
        \hline
        \textit{learning\_rate} & $0.001$ to $1$ & \textit{min\_child\_weight} & $1$ to $10$ \\
        \hline
        \textit{gamma} & $0.0001$ to $100$ & \textit{scale\_pos\_weight} & $\{1, 400/510\}$ \\
        \hline
    \end{tabular}
    \label{tab:microarrayHyperparameters}
    \end{minipage}
\end{table}

\begin{table}[h!]
    \centering 
    \caption{Single Cell Dataset Statistics}
    \label{table:TabellaDistribuzioneDatasetSingleCell}
    \begin{tabularx}{\textwidth}{l >{\centering\arraybackslash}X >{\centering\arraybackslash}X >{\centering\arraybackslash}X >{\centering\arraybackslash}X >{\centering\arraybackslash}X >{\centering\arraybackslash}X >{\centering\arraybackslash}X}
        \toprule 
        \textbf{Dataset} & \textbf{Cells} & \textbf{Genes} & \textbf{MS Patients} & \textbf{Control Patients} & \textbf{Patients} & \textbf{MS Cell Count} & \textbf{Control Cell Count} \\
        \midrule 
        CSF & 54,525 & 18,725 & 9 & 8 & 17 & 36,257 & 18,268 \\
        PBMC & 92,273 & 19,061 & 8 & 7 & 15 & 57,448 & 34,825 \\
        \bottomrule 
    \end{tabularx}
\end{table}

\begin{table}[h]
\centering
\caption{Number of clusters and average size for each dataset considered. We can observe how BCells tend to have significantly more clusters compared to CD4}
\begin{tabular}{lccc}
\textbf{Cluster} & \textbf{Total Clusters} & \textbf{Average Length} \\
\hline
CSF CD4 & 382 & 20 \\
CSF BCELLS & 4287 & 20 \\
PBMC BCELLS & 2422 & 70 \\
PBMC CD4 & 657 & 12 \\
\end{tabular}
\label{table:TabellaDistribuzioneClusterSingleCell}
\end{table}

\begin{table}[t]
    \footnotesize
    \centering
    \caption{Hyperparameters chosen for each model by the Bayesian Optimization applied during the search}
    \label{tab:ChoosenHyperparameters}

    \begin{tabular}{@{} l c c c c @{}} 
        \toprule
        \textbf{Hyperparameter} & \textbf{Microarray} & \textbf{CSF B-cells} & \textbf{CSF CD4} & \textbf{PBMC B-cells} \\
        \midrule
        \textbf{gamma} & 0.029 & 1e-06 & 0.0001 & 0.0001 \\
        \textbf{learning\_rate} & 0.107 & 0.050 & 0.166 & 0.084 \\
        \textbf{max\_depth} & 2 & 15 & 9 & 3 \\
        \textbf{min\_child\_weight} & 1 & 3 & 1 & 8 \\
        \textbf{n\_estimators} & 600 & 3000 & 600 & 1095 \\
        \textbf{reg\_alpha} & 0.008 & 0.0001 & 0.009 & 100.0 \\
        \textbf{reg\_lambda} & 0.0001 & 100.0 & 0.001 & 100.0 \\
        \textbf{scale\_pos\_weight} & 0.784 & 0.784 & 0.658 & 1.0 \\
        \bottomrule
    \end{tabular}
\end{table}

\begin{table}[t]
\footnotesize
 \centering
    \caption{Number of genes selected for each dataset by SHAP analysis.}
    \begin{tabular}{c|c|c|c}
    \textbf{Microarray} & \textbf{B cells CSF} & \textbf{CD4 CSF} & \textbf{B cells PBMC} \\
    \hline
        724 & 2115 & 1277 & 439 \\
    \hline
    \end{tabular}
    \label{tab:xAInumbers}
\end{table}


\begin{table}[h]
    \footnotesize
    \centering
    \caption{Top 10 enriched pathways for each dataset}
    
    \begin{minipage}{\textwidth}
        \centering
        \begin{tabular}{c|c}
            \textbf{Microarray} & \textbf{B cells CSF} \\
            \hline
            Disease & Eukaryotic Translation Elongation \\
            Immune System & Peptide chain elongation \\
            Cellular responses to stress & Viral mRNA Translation \\
            Cellular responses to stimuli & Formation of a pool of free 40S subunits \\
            Innate Immune System & Nonsense Mediated Decay (NMD) independent of the Exon Junction Complex (EJC) \\
            Generic Transcription Pathway & Selenocysteine synthesis \\
            RNA Polymerase II Transcription & Eukaryotic Translation Termination \\
            Infectious disease & L13a-mediated translational silencing of Ceruloplasmin expression \\
            Measles & GTP hydrolysis and joining of the 60S ribosomal subunit \\
            Cytokine Signaling in Immune system & Response of EIF2AK4 (GCN2) to amino acid deficiency \\
            \hline
        \end{tabular}
    \end{minipage}
    
    \vspace{0.5cm}
    
    \begin{minipage}{\textwidth}
        \centering
        \begin{tabular}{c|c}
            \textbf{CD4 CSF} & \textbf{B cells PBMC} \\
            \hline
            Eukaryotic Translation Elongation & Protein processing in endoplasmic reticulum \\
            Peptide chain elongation & Asparagine N-linked glycosylation \\
            Viral mRNA Translation & XBP1(S) activates chaperone genes \\
            Eukaryotic Translation Termination & Unfolded Protein Response (UPR) \\
            Nonsense Mediated Decay (NMD) independent of the Exon Junction Complex (EJC) & N-Glycan biosynthesis \\
            Selenocysteine synthesis & Various types of N-glycan biosynthesis \\
            Response of EIF2AK4 (GCN2) to amino acid deficiency & Metabolism of proteins \\
            Formation of a pool of free 40S subunits & Post-translational protein modification \\
            SRP-dependent cotranslational protein targeting to membrane & Transport to the Golgi and subsequent modification \\
            Nonsense Mediated Decay (NMD) enhanced by the Exon Junction Complex (EJC) & Cargo concentration in the ER \\
            \hline
        \end{tabular}
    \end{minipage}
    \label{tab:pathways}
\end{table}

\clearpage



\section*{Supplementary material}

**Complete SHAP and DEA results for the MSGD genes.**

| Gene | CSF_BCELLS | CSF_CD4 | PBMC B-cells | Microarray |
|---|---|---|---|---|
| APOA1 | None | Both | None | Not Found |
| OMG | SHAP | SHAP | None | None |
| CXCL8 | SHAP | SHAP | None | None |
| NEFL | None | SHAP | None | None |
| DDT | None | Not Found | None | None |
| ROCK1 | SHAP | SHAP | Not Found | None |
| KIR2DL3 | None | SHAP | DEA | Not Found |
| HBD | Not Found | Not Found | None | None |
| VIPR2 | SHAP | SHAP | None | Not Found |
| MANBA | None | None | None | None |
| SPG11 | None | None | None | None |
| SIRT1 | None | SHAP | None | None |
| HSP90B1 | None | None | Both | Not Found |
| MIR27B | Not Found | Not Found | Not Found | Not Found |
| MIR328 | Not Found | Not Found | Not Found | Not Found |
| IFI6 | None | None | None | None |
| PTPRC | None | SHAP | None | None |
| MIR18A | Not Found | Not Found | Not Found | Not Found |
| TOB1 | None | SHAP | DEA | None |
| HLA-DPB1 | Not Found | Both | None | None |
| MIR181A1 | Not Found | Not Found | Not Found | Not Found |
| MICB | None | SHAP | None | None |
| DCTN1 | None | SHAP | None | None |
| RAC2 | None | SHAP | DEA | None |
| HAR1B | SHAP | None | None | Not Found |
| MT-ND6 | None | SHAP | None | Not Found |
| BDNF-AS | SHAP | SHAP | None | Not Found |
| TNF | None | None | DEA | None |
| MIF | None | SHAP | None | Not Found |
| GFI1 | DEA | None | None | DEA |
| MTHFR | SHAP | SHAP | None | Not Found |
| CD4 | SHAP | SHAP | None | None |
| DGCR8 | None | SHAP | None | None |
| TET2 | None | SHAP | None | Not Found |
| TIMP1 | None | SHAP | DEA | None |
| RNF213 | None | None | None | Not Found |
| TCF4 | None | Both | None | None |
| EIF4EBP2 | None | None | None | Not Found |
| APAF1 | Not Found | None | None | None |
| MIR326 | Not Found | Not Found | Not Found | Not Found |
| PSMB8 | None | SHAP | None | None |
| CD1E | Not Found | SHAP | DEA | None |
| IL1B | Not Found | SHAP | None | SHAP |
| PLXNC1 | None | SHAP | Not Found | Not Found |
| IFNAR2 | None | None | DEA | None |

| Gene | Col2 | Col3 | Col4 | Col5 |
|------|------|------|------|------|
| FLT1 | Not Found | SHAP | None | Not Found |
| MIR193A | Not Found | Not Found | Not Found | Not Found |
| CD58 | None | None | DEA | None |
| PDXP | Both | Not Found | DEA | Not Found |
| ETS1 | SHAP | None | None | None |
| ADRB3 | Not Found | Not Found | Not Found | Not Found |
| TUBA1C | Not Found | SHAP | None | Not Found |
| NFKB1 | Not Found | SHAP | None | None |
| HLA-DRA | Not Found | Both | None | None |
| NAMPT | None | None | None | None |
| PDCD1 | None | SHAP | None | Not Found |
| PDCD1LG2 | None | Both | None | Not Found |
| MIR16-1 | Not Found | Not Found | Not Found | Not Found |
| ATG16L2 | None | SHAP | None | Not Found |
| MIR19A | Not Found | Not Found | Not Found | Not Found |
| EIF2B5 | None | SHAP | None | None |
| IFNAR1 | None | None | None | None |
| RUNX1 | None | SHAP | None | None |
| ARG1 | None | Both | DEA | Not Found |
| MIR374B | Not Found | Not Found | Not Found | Not Found |
| CD14 | None | SHAP | None | None |
| RHOA | SHAP | Not Found | None | None |
| SERTAD1 | None | SHAP | None | Not Found |
| CYBB | SHAP | SHAP | None | None |
| NT5E | Not Found | SHAP | None | None |
| CXCL12 | None | SHAP | None | Not Found |
| ICAM1 | None | SHAP | DEA | None |
| CAT | None | SHAP | None | Not Found |
| DRD3 | SHAP | SHAP | DEA | None |
| NEAT1 | SHAP | None | DEA | Not Found |
| LASP1 | SHAP | SHAP | None | None |
| IFIH1 | None | None | None | None |
| NFKBID | None | SHAP | None | Not Found |
| IL7R | SHAP | None | DEA | Not Found |
| HLA-DRB2 | Not Found | Not Found | Not Found | Not Found |
| LY86 | None | Not Found | Not Found | None |
| IL-35 | Not Found | Not Found | Not Found | Not Found |
| SLC11A1 | None | Both | None | None |
| CXCR4 | None | None | DEA | None |
| IL27 | Not Found | SHAP | None | Not Found |
| MIR454 | Not Found | Not Found | Not Found | Not Found |
| NR1I2 | None | SHAP | None | Not Found |
| DROSHA | SHAP | SHAP | None | None |
| BB | Not Found | Not Found | Not Found | Not Found |
| MIR146A | Not Found | Not Found | Not Found | Not Found |
| MIR137 | Not Found | Not Found | Not Found | Not Found |

| Gene | Col2 | Col3 | Col4 | Col5 |
|---|---|---|---|---|
| REG1A | Not Found | Not Found | None | Not Found |
| MIR181C | Not Found | Not Found | Not Found | Not Found |
| LIF | DEA | Both | None | Not Found |
| IL17 | Not Found | Not Found | Not Found | Not Found |
| CD274 | None | None | None | Not Found |
| CLU | SHAP | Both | Not Found | Not Found |
| MIR128 | Not Found | Not Found | Not Found | Not Found |
| IL7 | SHAP | None | Not Found | None |
| CXCR3 | SHAP | None | DEA | None |
| MIR106A | Not Found | Not Found | Not Found | Not Found |
| HTN1 | SHAP | SHAP | None | Not Found |
| SIRPG | None | None | None | None |
| CYP27B1 | None | Both | None | Not Found |
| MIR484 | Not Found | Not Found | Not Found | Not Found |
| PINK1-AS | None | SHAP | None | Not Found |
| CRB3 | None | None | None | Not Found |
| THRIL | None | SHAP | None | Not Found |
| CXCL10 | None | SHAP | None | None |
| HLA-B | Not Found | Not Found | Not Found | Not Found |
| A2M | SHAP | DEA | None | Not Found |
| RAC1 | Not Found | None | None | Not Found |
| PSEN1 | SHAP | None | None | Not Found |
| MIR340 | Not Found | Not Found | Not Found | Not Found |
| IL21 | None | Both | DEA | Not Found |
| IL17A | None | None | None | Not Found |
| VIPR1 | None | SHAP | None | Not Found |
| NFATc2 | Not Found | Not Found | Not Found | Not Found |
| SGPL1 | Not Found | SHAP | None | None |
| HLA-DRB1 | Not Found | Not Found | None | Both |
| MIR922 | Not Found | Not Found | Not Found | Not Found |
| CD44 | None | None | None | None |
| STAT1 | None | SHAP | None | None |
| MIRLET7E | Not Found | Not Found | Not Found | Not Found |
| BATF | None | None | DEA | None |
| CYP24A1 | None | Both | None | Not Found |
| MYC | None | None | None | Not Found |
| HLA-DPA1 | SHAP | Both | None | None |
| TAGAP | None | SHAP | None | Not Found |
| CD226 | None | None | None | Not Found |
| SERPINE1 | None | SHAP | None | Not Found |
| CDKN1A | SHAP | None | DEA | None |
| BDKRB1 | Both | SHAP | DEA | Not Found |
| IL6ST | SHAP | SHAP | None | None |
| FOXP3 | SHAP | SHAP | None | Not Found |
| CDKN2B-AS1 | None | SHAP | None | Not Found |
| APOE | None | Both | None | Not Found |

| Gene | Col2 | Col3 | Col4 | Col5 |
|---|---|---|---|---|
| MIR125B1 | Not Found | Not Found | Not Found | Not Found |
| TUG1 | SHAP | None | None | Not Found |
| IL10 | None | SHAP | DEA | Not Found |
| MT-ATP6 | Not Found | SHAP | Not Found | Not Found |
| EVI5 | SHAP | SHAP | None | None |
| CCR8 | None | SHAP | None | Not Found |
| CD24 | SHAP | SHAP | None | None |
| ATG5 | Not Found | SHAP | None | Not Found |
| CCR4 | None | None | None | Not Found |
| IKZF2 | SHAP | SHAP | None | Not Found |
| CX3CR1 | None | None | None | None |
| CBLB | None | None | None | None |
| YAP1 | SHAP | None | DEA | Not Found |
| MMP2 | None | SHAP | Not Found | Not Found |
| DRD5 | Not Found | Not Found | Not Found | Not Found |
| PTPN6 | None | None | DEA | None |
| MIR29B1 | Not Found | Not Found | Not Found | Not Found |
| PRF1 | Both | SHAP | Not Found | None |
| WWTR1 | SHAP | SHAP | None | Not Found |
| BCL2 | None | None | None | None |
| RMRP | None | None | Not Found | Not Found |
| ADRB1 | None | SHAP | None | Not Found |
| C4A | None | SHAP | DEA | Not Found |
| MIR26A1 | Not Found | Not Found | Not Found | Not Found |
| ERVW-1 | SHAP | SHAP | DEA | Not Found |
| HLA-DRB4 | Not Found | Not Found | Not Found | Not Found |
| NGF | None | SHAP | Not Found | Not Found |
| TNFSF14 | Not Found | None | None | None |
| CR2 | Not Found | SHAP | DEA | None |
| MIRLET7F1 | Not Found | Not Found | Not Found | Not Found |
| YTHDF1 | None | SHAP | None | None |
| EXTL3 | Not Found | SHAP | None | None |
| MAGI2-AS3 | Not Found | Not Found | Not Found | Not Found |
| ZMIZ1 | Not Found | None | DEA | None |
| MPO | None | SHAP | None | None |
| KL | None | SHAP | None | Not Found |
| SOCS1 | None | SHAP | None | None |
| IL2RA | DEA | None | DEA | SHAP |
| SIL1 | None | SHAP | SHAP | None |
| FCRL2 | Not Found | SHAP | None | Not Found |
| CD86 | None | Both | DEA | None |
| miR-155 | Not Found | Not Found | Not Found | Not Found |
| HSPA5 | None | Not Found | None | None |
| MIR140 | Not Found | Not Found | Not Found | Not Found |
| MIR20A | Not Found | Not Found | Not Found | Not Found |
| POLR1D | SHAP | Not Found | None | None |

| Gene | Col2 | Col3 | Col4 | Col5 |
|---|---|---|---|---|
| TALDO1 | None | Not Found | DEA | Not Found |
| ADRA2C | Not Found | Not Found | DEA | Not Found |
| IL15 | Not Found | SHAP | None | None |
| APP | None | SHAP | None | Not Found |
| YTHDF2 | SHAP | SHAP | None | None |
| RGCC | None | SHAP | DEA | None |
| SPP1 | SHAP | Not Found | None | Not Found |
| PLP1 | SHAP | SHAP | None | None |
| TRADD | Not Found | Not Found | DEA | None |
| USP18 | None | SHAP | None | None |
| MMP9 | None | SHAP | None | None |
| HSPA1B | None | SHAP | DEA | Not Found |
| TIGIT | Both | SHAP | None | Not Found |
| CYP4F2 | None | SHAP | DEA | Not Found |
| ERG | None | SHAP | Not Found | Not Found |
| ACTB | None | SHAP | None | None |
| CPSF7 | Not Found | SHAP | None | Not Found |
| TYK2 | None | None | None | None |
| MIR92A1 | Not Found | Not Found | Not Found | Not Found |
| IL23A | None | None | None | None |
| IRF1 | None | SHAP | None | None |
| TBX21 | None | SHAP | Not Found | None |
| ITGAM | SHAP | SHAP | DEA | None |
| STAT3 | SHAP | None | None | None |
| IL18 | SHAP | Both | Not Found | None |
| MIR145 | Not Found | Not Found | Not Found | Not Found |
| FCRL5 | Not Found | SHAP | None | Not Found |
| HAVCR2 | SHAP | SHAP | None | Not Found |
| PANDAR | None | SHAP | None | Not Found |
| ITGB1 | None | None | DEA | None |
| IFNA2 | Not Found | Not Found | Not Found | Not Found |
| VRK2 | None | None | None | None |
| DRD2 | Not Found | Not Found | Not Found | Not Found |
| SOCS5 | None | SHAP | None | None |
| STAT5A | None | SHAP | None | None |
| CCR1 | SHAP | Both | None | None |
| TNFRSF11B | SHAP | Not Found | None | Not Found |
| CX3CL1 | None | SHAP | None | Not Found |
| HAR1A | None | SHAP | None | Not Found |
| ATF4 | None | None | None | None |
| CNR1 | None | Both | None | Not Found |
| PRKCA | None | SHAP | Not Found | None |
| HHEX | None | SHAP | None | None |
| ADRA2A | Not Found | Not Found | None | None |
| TGFB2 | None | SHAP | None | Not Found |
| TLR9 | SHAP | None | DEA | Not Found |

| Gene | Col2 | Col3 | Col4 | Col5 |
|---|---|---|---|---|
| CNR2 | Not Found | Both | None | Not Found |
| ZFP36L1 | SHAP | None | None | None |
| IL12RB2 | None | None | None | Not Found |
| JUN | None | None | None | Not Found |
| GPR183 | None | SHAP | DEA | None |
| LTA | Both | None | None | Not Found |
| RGS1 | SHAP | SHAP | None | None |
| BACH2 | None | SHAP | Both | None |
| ABCB1 | None | SHAP | DEA | Not Found |
| ADRA2B | SHAP | SHAP | None | None |
| MIR17 | Not Found | Not Found | Not Found | Not Found |
| DDIT3 | None | SHAP | None | None |
| TLR2 | SHAP | Both | None | None |
| MIR633 | Not Found | Not Found | Not Found | Not Found |
| NR4A2 | SHAP | SHAP | None | None |
| TNFSF15 | None | SHAP | None | None |
| IL2 | DEA | SHAP | DEA | Not Found |
| MAG | None | Both | None | Not Found |
| WTAP | None | None | None | None |
| CD1A | None | SHAP | DEA | None |
| TNFSF13B | None | None | None | Not Found |
| TIAM1 | SHAP | None | None | None |
| TGFB1 | SHAP | None | None | Not Found |
| RTL1 | Not Found | Not Found | Not Found | Not Found |
| CLDN11 | None | Not Found | Not Found | Not Found |
| IFNG | DEA | None | None | Not Found |
| FCRL1 | Not Found | None | Both | Not Found |
| KLRB1 | SHAP | SHAP | None | Not Found |
| MBP | None | SHAP | None | DEA |
| ADRB2 | None | None | DEA | None |
| NTSR1 | DEA | SHAP | None | None |
| DICER1 | SHAP | None | Not Found | None |
| GAPDH | None | None | DEA | Not Found |
| MIR330 | Not Found | Not Found | Not Found | Not Found |
| IRF7 | None | SHAP | None | None |
| DDX5 | None | SHAP | None | None |
| BDNF | SHAP | SHAP | None | Not Found |
| LRPPRC | Not Found | SHAP | None | None |
| ABCG2 | Not Found | SHAP | None | Not Found |
| STAT6 | SHAP | SHAP | None | None |
| AIM2 | Both | SHAP | DEA | None |
| IL1R1 | None | Both | DEA | Not Found |
| LILRA3 | Not Found | Not Found | Not Found | None |
| STAT4 | None | SHAP | None | None |
| CD40 | None | SHAP | Not Found | None |
| NINJ2 | SHAP | None | Not Found | None |

| Gene | Col1 | Col2 | Col3 | Col4 |
|---|---|---|---|---|
| MAP2 | None | SHAP | DEA | Not Found |
| TRAF2 | None | SHAP | DEA | None |
| ODAD1 | Not Found | Not Found | Not Found | Not Found |
| MIRLET7F2 | Not Found | Not Found | Not Found | Not Found |
| OIP5‚ÄêAS1 | Not Found | Not Found | Not Found | Not Found |
| MEFV | None | SHAP | None | None |
| MT-CYB | Not Found | SHAP | Not Found | Not Found |
| MIR15A | Not Found | Not Found | Not Found | Not Found |
| GAPVD1 | Not Found | SHAP | None | None |
| IFNGR1 | None | SHAP | None | None |
| PLEK | SHAP | SHAP | None | None |
| CCL3 | SHAP | Both | None | Not Found |
| IDO1 | None | SHAP | None | Not Found |
| TNFSF4 | Not Found | SHAP | DEA | None |
| P2RY12 | Not Found | SHAP | None | Not Found |
| IL-10 | Not Found | Not Found | Not Found | Not Found |
| MAPT | None | SHAP | None | Not Found |
| CCL4 | None | Not Found | None | None |
| MIR1-1 | Not Found | Not Found | Not Found | Not Found |
| PDCD5 | SHAP | None | None | Not Found |
| MIR548AC | Not Found | Not Found | Not Found | Not Found |
| CCL5 | SHAP | None | None | None |
| SP3 | None | SHAP | Not Found | None |
| RORC | SHAP | Not Found | DEA | None |
| C4B | Not Found | Not Found | DEA | Not Found |
| LTB | None | None | None | Not Found |
| RIGI | Not Found | Not Found | Not Found | None |
| MX1 | None | SHAP | None | None |
| IKBKE | None | SHAP | None | None |
| PTEN | None | None | None | None |
| AVP | None | SHAP | Not Found | None |
| NFE2L2 | None | None | None | None |
| SEMA3A | DEA | DEA | None | Not Found |
| ALKBH5 | None | SHAP | None | Not Found |
| VIRMA | Not Found | Not Found | Not Found | Not Found |
| S100A6 | None | Not Found | DEA | None |
| SELPLG | None | SHAP | None | Not Found |
| CD8A | Both | Not Found | None | None |
| CSTF2 | None | None | None | None |
| ADRA1B | SHAP | Not Found | Not Found | Not Found |
| GAS5 | Not Found | Not Found | Not Found | Not Found |
| NRP1 | SHAP | SHAP | Not Found | Not Found |
| FADD | None | SHAP | None | None |
| ANKRD55 | None | SHAP | DEA | Not Found |
| LAG3 | Not Found | SHAP | None | None |
| IL6 | None | Both | Not Found | Not Found |

| Gene | Col2 | Col3 | Col4 | Col5 |
|---|---|---|---|---|
| IGF1 | SHAP | Both | Not Found | Not Found |
| NLRP3 | SHAP | SHAP | None | None |
| MMP-2 | Not Found | Not Found | Not Found | Not Found |
| CD40LG | None | SHAP | None | None |
| Fasl | Not Found | Not Found | Not Found | Not Found |
| ADRA1A | None | SHAP | None | Not Found |
| TNFRSF25 | None | None | DEA | Not Found |
| CD22 | SHAP | SHAP | DEA | None |
| RRN3 | None | SHAP | None | None |
| CASP1 | None | None | DEA | Not Found |
| MIR497 | Not Found | Not Found | Not Found | Not Found |
| IL16 | SHAP | SHAP | None | None |
| HNRNPC | None | None | Not Found | Not Found |
| MIR126 | Not Found | Not Found | Not Found | Not Found |
| CTLA4 | Not Found | SHAP | None | None |
| DDX39B | None | SHAP | None | Not Found |
| RORA | None | None | DEA | Not Found |
| NDFIP1 | None | None | DEA | None |
| CCR5 | None | None | None | Not Found |
| NLRC4 | None | Both | None | Not Found |
| VEGFA | SHAP | SHAP | DEA | Not Found |
| METTL3 | None | None | None | None |
| IL1RN | None | SHAP | None | None |
| HLA-C | Not Found | None | Not Found | Not Found |
| HLA-DRB3 | Not Found | Not Found | Not Found | Not Found |
| MT-CO2 | Not Found | Not Found | Not Found | Not Found |
| MIR155 | Not Found | Not Found | Not Found | Not Found |
| AR | DEA | SHAP | Not Found | None |
| EOMES | Both | Not Found | DEA | Not Found |
| TLR4 | SHAP | Both | None | None |
| GATA3 | Not Found | None | DEA | Not Found |
| VDR | None | SHAP | Not Found | Not Found |
| HLA-DRB5 | SHAP | SHAP | Not Found | Both |
| HSPA4 | None | SHAP | None | None |
| MIR590 | Not Found | Not Found | Not Found | Not Found |
| ERMN | None | SHAP | None | Not Found |
| BIRC5 | SHAP | Not Found | None | Not Found |
| TNFSF10 | None | SHAP | DEA | None |
| CXCL13 | Both | None | DEA | Not Found |
| SOD1 | None | Not Found | None | None |
| CCR2 | None | SHAP | DEA | None |
| SEMA7A | None | SHAP | Not Found | Not Found |
| ADRA1D | Not Found | Not Found | None | Not Found |
| CFH | None | SHAP | DEA | None |
| STAT2 | SHAP | SHAP | None | None |
| CCL2 | None | SHAP | None | None |

| | | | | |
|---|---|---|---|---|
| HOTAIR | Not Found | Not Found | Not Found | Not Found |
| COX5B | None | SHAP | None | Not Found |
| GSDMB | None | None | None | None |
| PTGER4 | None | None | None | None |
| CLEC16A | None | SHAP | None | Not Found |
| GZMB | Both | Both | None | Not Found |
| WFDC21P | Not Found | Not Found | Not Found | Not Found |
| MALAT1 | Both | SHAP | None | Not Found |
| TGFB3 | SHAP | SHAP | None | None |
| TP53 | SHAP | SHAP | None | None |
| DNMT1 | None | None | None | None |
| IL12A | None | SHAP | None | None |
| IL12B | Not Found | SHAP | None | Not Found |
| PLAU | None | SHAP | None | Not Found |
| NRON | Not Found | Not Found | Not Found | Not Found |
| TYRO3 | None | SHAP | DEA | Not Found |



\end{document}